\documentclass[twocolumn,showpacs,preprintnumbers,amsmath,amssymb]{revtex4}
\usepackage{graphicx}% Include figure files
\usepackage{dcolumn}% Align table columns on decimal point
\usepackage{bm}% bold math

\begin{document}

\preprint{APS/123-QED}

%\title{The discharge of a granular silo from discrete to continuum}% Force line breaks with \\

\title{Continuum simulation of the discharge of the granular silo:\\ a validation test for the $\mu(I)$-visco-plastic flow law}

\author{L. Staron$^{1,2}$, P.-Y. Lagr\'ee$^2$ and S. Popinet$^3$}
%\email{lydie.staron@upmc.fr}

\affiliation{%
$^1$ School of Earth Sciences, University of Bristol, Queens Road, Bristol BR8 1RJ, United Kingdom.\\
$^2$CNRS - Universit\'e Pierre et Marie Curie Paris 6, UMR 7190, Institut Jean Le Rond d'Alembert, F-75005 Paris, France.\\
$^3$National Institute of Water and Atmospheric Research, PO Box 14-901 Kilbirnie, Wellington, New Zealand.
}%

\date{\today}% It is always \today, today,

\begin{abstract}
Using both a continuum Navier-Stokes solver, with the $\mu(I)$-flow-law implemented to model the viscous behavior, and the discrete Contact Dynamics algorithm, 
 the discharge of granular silos is simulated in two dimensions from the early stages of the discharge until complete release of the material. In both cases, the Beverloo scaling is recovered. We first do not attempt quantitative comparison, but focus on the qualitative behavior of velocity and pressure at different locations in the flow. A good agreement is obtained in the regions of rapid flows, while areas of slow creep are not entirely captured by the continuum model. The pressure field shows a general good agreement. The evolution of the free surface implies differences, however, the bulk deformation is essentially identical in both approaches. 
 The influence of the parameters of the $\mu(I)$-flow-law is systematically investigated,  showing the importance of the dependence on the inertial number $I$ to achieve quantitative agreement between continuum and discrete discharge. The general ability of the continuum model to reproduce qualitatively the granular behavior is found to be very encouraging.
\end{abstract}

\maketitle

\section{Introduction}

\begin{figure*}
\begin{minipage}{0.60\linewidth} 
\begin{minipage}{0.45\linewidth} 
\centerline{\includegraphics[angle = -90,width = 1.\linewidth]{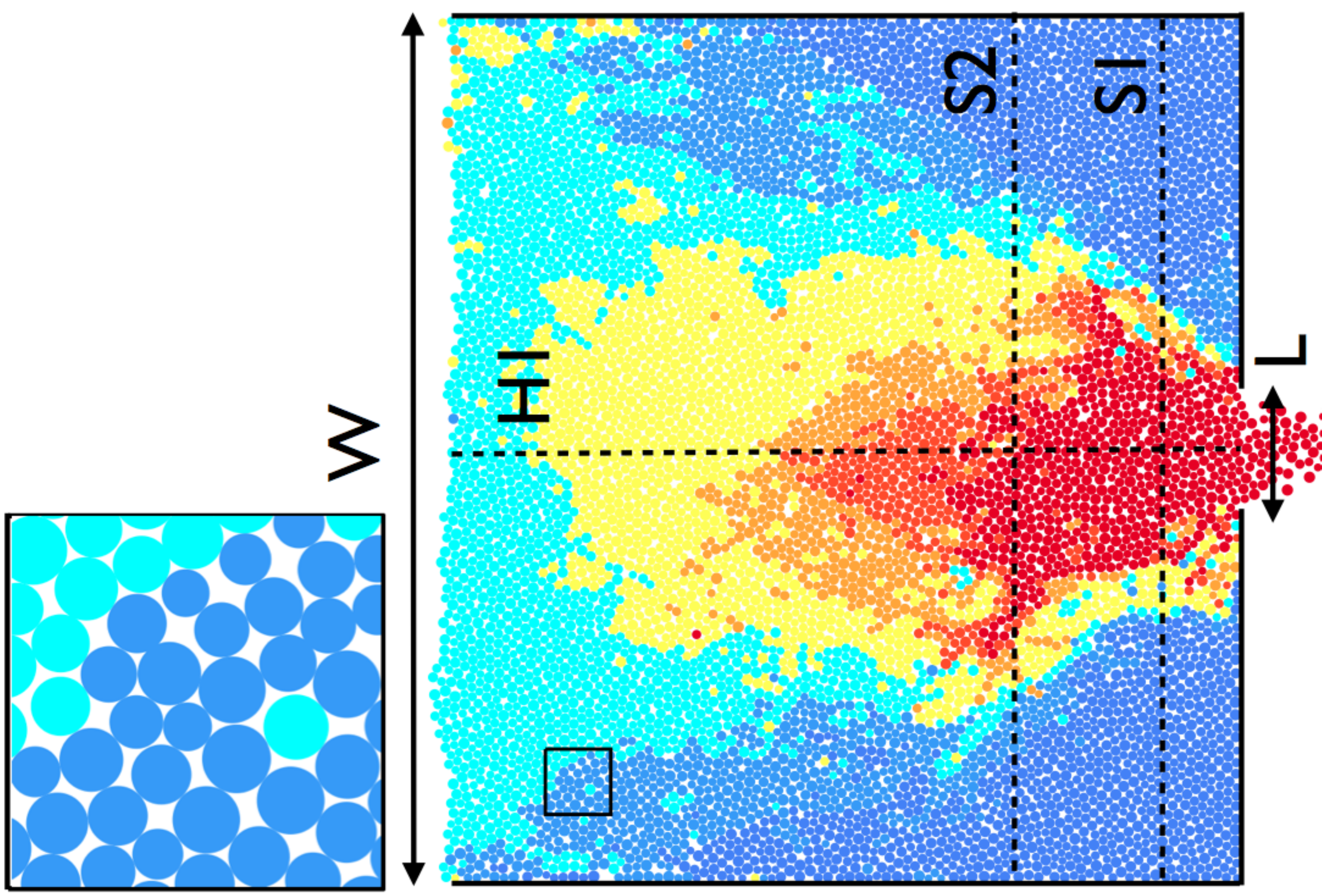}}
\end{minipage}
\hfill
\begin{minipage}{0.45\linewidth} 
\centerline{\includegraphics[angle = -90,width = 1.\linewidth]{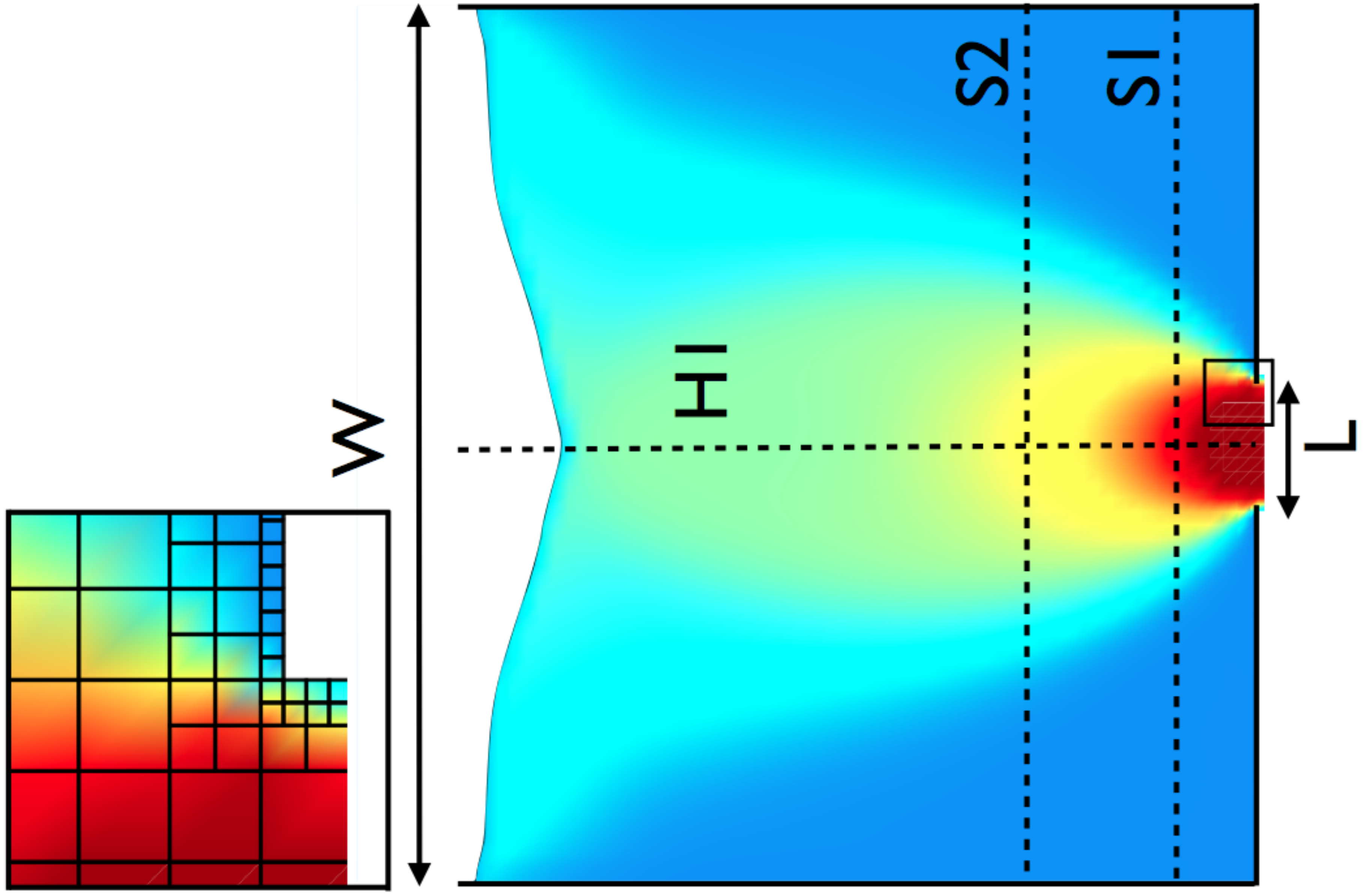}}
\end{minipage}
\caption{(Color On-line) Discrete silo simulated by Contact Dynamics (left) and continuum silo simulated by Gerris (right); the outlet size is $L$, the silo width is $W$; $S_1$, $S_2$ and $H_1$ are cross-sections along which velocity and pressure are analyzed.} 
\label{illus}
\end{minipage}
\end{figure*}

Granular matter is a well-known example of complex material able to flow like a viscous fluid or resist shear stress like a solid, and evolving from one state to the other over a distance of typically  few grain diameters.  During the discharge of a silo, this property is responsible for the coexistence of rapid dilute flow in the vicinity of  the outlet, dense slower shear in the higher parts of the bulk, and static regions at the bottom of the container. In some instances, when the outlet can accommodate only few particles diameters, arching occurs, that is the formation of highly loaded force chains above the orifice, whereby flow is stopped, or made intermittent \cite{lepennec98,janda09}.  \\
Silos are widely used in geo-technical or agro-technical applications, for which the full understanding of the discharge dynamics and its reliable modeling are critical \cite{walker66,davies83}.  Meanwhile, the variety of behaviors exhibited in a silo justifies the large academic interest granted to the subject. As a result, much understanding has been gained on the silo phenomenology, from "why hour glasses tick", to the Beverloo scaling for the discharge rate or the shape of the free surface \cite{beverloo61,wu93,gray97,samadani02}. Because of the specificity of its behavior, the granular silo is a stringent test for continuum modeling of granular matter \cite{rotter98,kamrin10,staron12,sun12}. The fact that the silo outlet may be of a little number of grains size, hence threatening the validity of a continuous approach, forms a first complication: indeed, uninterrupted flows may be obtained for outlet size as small as 5 grain diameters \cite{mankoc07}. While the modeling of the intermittent flow regime is hardly accessible to continuum modeling, it is not clear that well-developed flows over such small length-scales are well captured either \cite{rycroft09}.   
The main difficulty however is the simultaneous existence of static and rapidly flowing zones, which requires a unified picture of what is often described as solid-like and fluid-like behaviors, each of them forming a challenge of its own. The solid-like behavior of granular matter is characterized by the small domain of elastic response, a plastic threshold whose dependence on the grains properties and packing history is unclear, and important force fluctuations which may compromise the validity of a continuum picture at the scale of few grain diameters \cite{claudin98,roux00}.  
The fluid-like behavior also has its share of difficulty, and offers a wide variety of complicated behaviors depending on the system geometry, which have recently  benefited from important progress with the formulation of the $\mu(I)$-flow-law \cite{gdrmidi04,dacruz05,jop06}. 
Achieving a continuum picture of a system as complete as the granular silos requires a reliable physical modeling of both fluid-like and solid-like properties. This may be undertaken either by generalizing elasto-plastic approaches to rapidly moving zone \cite{rotter98,crosta09}, or by considering the system as a viscous flow with areas of infinite viscosity \cite{lagree11, staron12}.   The first approach was applied in \cite{kamrin10} to the initial stages of the silo discharge, where stress and velocity fields were found to match those observed in DEM simulations. Therefore, the plastic part of the deformations (namely developed flow)  was chosen to obey the $\mu(I)$-flow-law.  In this contribution, we adopt the fluid approach, namely we approximate the granular matter as a viscous material flowing following fluid mechanics equations. As in \cite{jop06,lagree11,staron12}, and following the choice of \cite{kamrin10}, we adopt the $\mu(I)$-flow-law to describe the viscous behavior of the granular matter \cite{gdrmidi04}. Doing so, we are able to simulate in two dimensions all the stages of the silo discharge, from onset to the complete release of the material. Comparison with discrete simulations of granular silos are carried out. In both cases, the Beverloo scaling is recovered. Adjusting rheological parameters to make discrete and continuum discharge coinciding being non-trivial, we first do not attempt quantitative comparison, but focus on the qualitative behavior of velocity and pressure at different locations in the flow. A good agreement is obtained in the regions of rapid flows, while areas of slow creep are not entirely captured by the continuum model.  The pressure field shows a general good agreement. The evolution of the free surface also exhibits differences, however, the bulk deformation shows a good agreement between the two approaches. 
 The influence of the parameters of the $\mu(I)$-flow-law is systematically investigated,  showing the importance of the dependence on the inertial number $I$ to achieve quantitative agreement between continuum and discrete discharge, and allowing discussion on the limitations of the model.  Finally, the role of the boundary conditions is studied in a last section.

%-----------------------------------------
\section{Modeling discrete and continuum granular silos}

\subsection{A continuum model for granular flows}
\label{sec:modA}
Defining mean viscous properties for granular flows that would be able to describe the sharp transition between rapid flow, creep motion and quasi-static state (observed for instance  in the silo configuration, in avalanches on erodible beds, and in all transient flows) has proven a long-lasting obstacle to efficient modeling of granular flows. The most obvious difficulty is to characterize the divergence of the viscosity, or jamming transition. A strategy to bypass this difficulty is to forego the explicit definition of the viscosity and to rely instead on the frictional properties of granular matter which relate pressure and shear stress: $\tau= \mu P$, where $\mu$ is the effective friction of the material and $P$ the pressure. Assuming that shear rate and shear deformation are collinear (which is the case in simple flow configuration like chute flows, but less obvious in more complex configuration like silos of collapsing columns \cite{lacaze09,lagree11}), a relation can be derived between shear rate and shear stress which can be used as a substitute for viscosity:
\begin{equation}
\bold{\tau} = \frac{\mu P}{\| \dot{\gamma} \|}  \dot{\gamma},
\label{eq:eta}
\end{equation}
where $ \dot{\gamma}$ and $\| \dot{\gamma} \|$ are the shear rate and the norm of the shear rate respectively. This simplification was first successfully applied for a simple flow configuration, and later in more complex configurations, including the silo \cite{jop06,chauchat10,kamrin10,lagree11,staron12}. This strategy is also adopted in this work. \\
According to relation (\ref{eq:eta}), we see that a constant friction model ($\mu=$ cst), as simple as it is, will nevertheless lead to a non trivial viscous behavior, showing shear-thinning properties and a dependence on the local pressure. This case is addressed in section~\ref{tunemui}. In this contribution however, we are interested is assessing the performances of the $\mu(I)$-flow law. Established on the basis of experimental and numerical works in various simple flow configurations (planar shear, couette flow, chute flows and rotating drums \cite[and reference therein]{gdrmidi04}),  it has since led to the successful recovery of granular dynamics in more testing situations: 3D chute flow with rough side-walls \cite{jop06}, the early stage of the discharge of a granular silo \cite{kamrin10}, or the collapse of granular columns under gravity \cite{lagree11} for instance.\\
The  $\mu(I)$-flow-law implemented in this work is identical to that used in  \cite{jop06}: $\mu$ is a function of the non-dimensional number $I = d \| \dot{\gamma} \| /\sqrt{P/\rho}$, where $d$ is the mean grain diameter and $\rho$ the density,  following the dependence
\begin{equation}
\mu = \mu_s + \frac{\Delta \mu} {1+ I_0/I },
\label{muI}
\end{equation}
 where $\mu_s$, $\Delta \mu$ and $I_0$ are constants \cite{jop06}.
 Based on previous work comparing the rapid flow of discrete granular systems and their continuum counterpart in the column collapse configuration \cite{lagree11}, we first chose  $\mu_s = 0.32$,  $\Delta \mu=0.28$ and $I_0=0.4$. The influence of the value of these parameters is specifically addressed in section \ref{tunemui}.   \\
  The dependence of the friction properties on the non-dimensional number $I$  (whose relevance to granular flows was also discussed in \cite{ancey99}) conveys the fact that the local dynamics of the grains rearranging under a given pressure when submitted to a given macroscopic deformation reflects in the dissipation properties. It describes a dependence on the dynamics, according which the frictional properties of the flow vary between two extremal values: a smaller one corresponding to static state and a larger one corresponding to rapid flow.   
  The precise shape of the dependence itself, as observed in experiments and simulations, may be questioned: in \cite{hatano07} for instance, a power-law dependence is proposed. In this contribution, the sensitivity of the results to the shape of the  $\mu(I)$-law (equation (\ref{muI})) is not investigated. However, we discuss in detail the implication of the $I$-dependence itself, and its role in the ability of the continuum model to reproduce the outcome of discrete simulations (section~\ref{tunemui}).\\
 More elaborate rheological models may account for non-local effects \cite{pouliquen09,kamrin12} or incorporate explicitly the granular micro-structure \cite{sun11}. These aspects are not included in the present discussion. We will see however that, in spite of its simplicity, the $\mu(I)$-flow-law leads to the recovery of a large part of the granular silo phenomenology, which we reproduce using discrete numerical simulations.

%---------- BEVERLOO---------------
\begin{figure*}
\begin{minipage}[t]{0.48\linewidth} 
\centerline{\includegraphics[width = 0.9\linewidth]{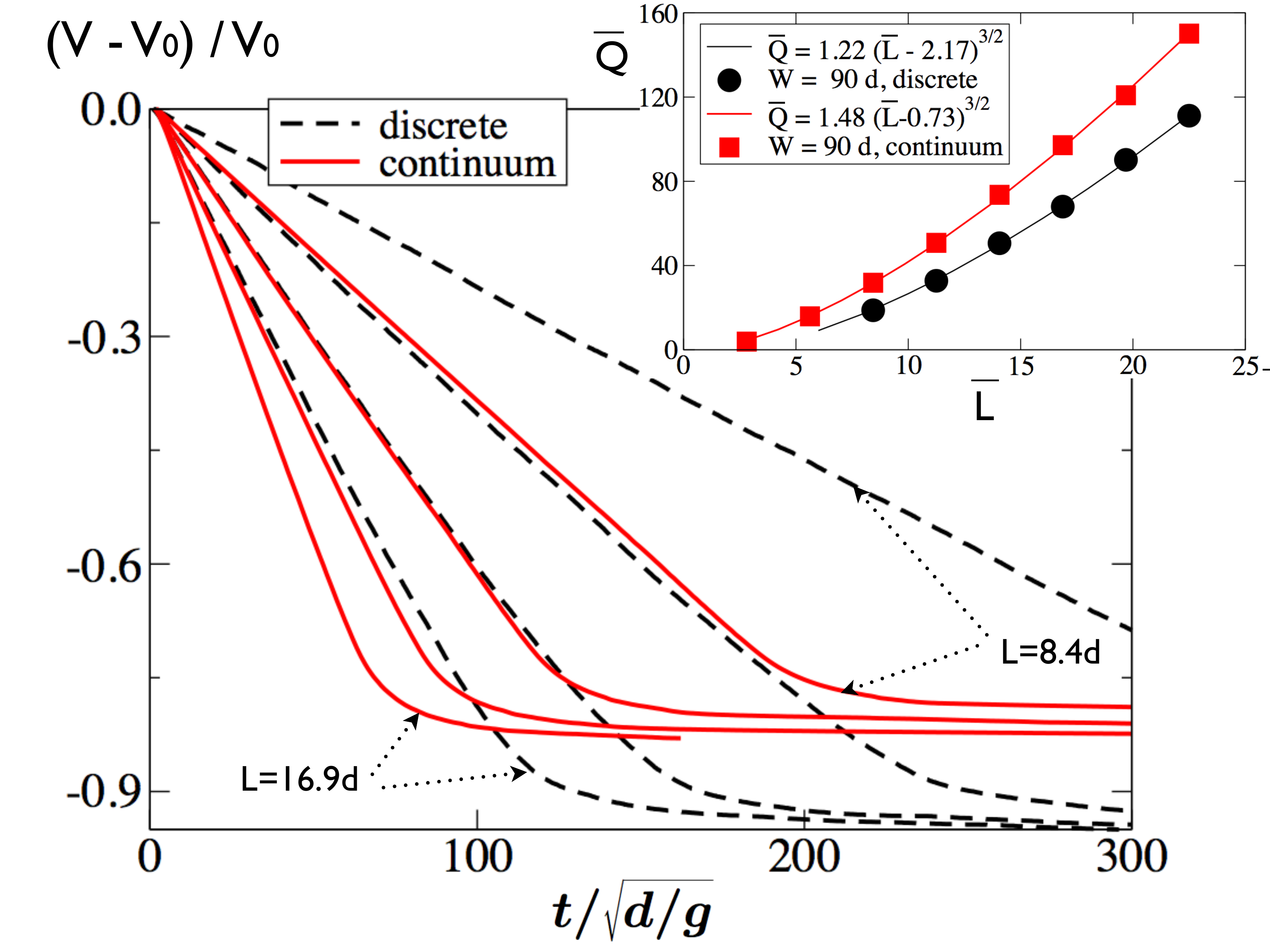}}
\caption{Normalized volume of granular matter left in the silo as a function of the normalized time $t/\sqrt{d/g}$  for discrete (dashed-line) and continuum (plain line) simulations, for outlet $L/d= 8.4$, $11.2$, $14.1$ and $16.9$. Inset: Normalized discharge rate $\bar{Q} = Q/\sqrt{g}d^{3/2}$ as a function of the normalized outlet size  $\bar{L} = L/d$  with corresponding Beverloo scalings  (see equation (\ref{eq:beverloo})). }
\label{Discharge}
\end{minipage}
\hfill
%\end{figure}
%
%\begin{figure}
\begin{minipage}[t]{0.48\linewidth} 
\centerline{\includegraphics[width = 0.9\linewidth]{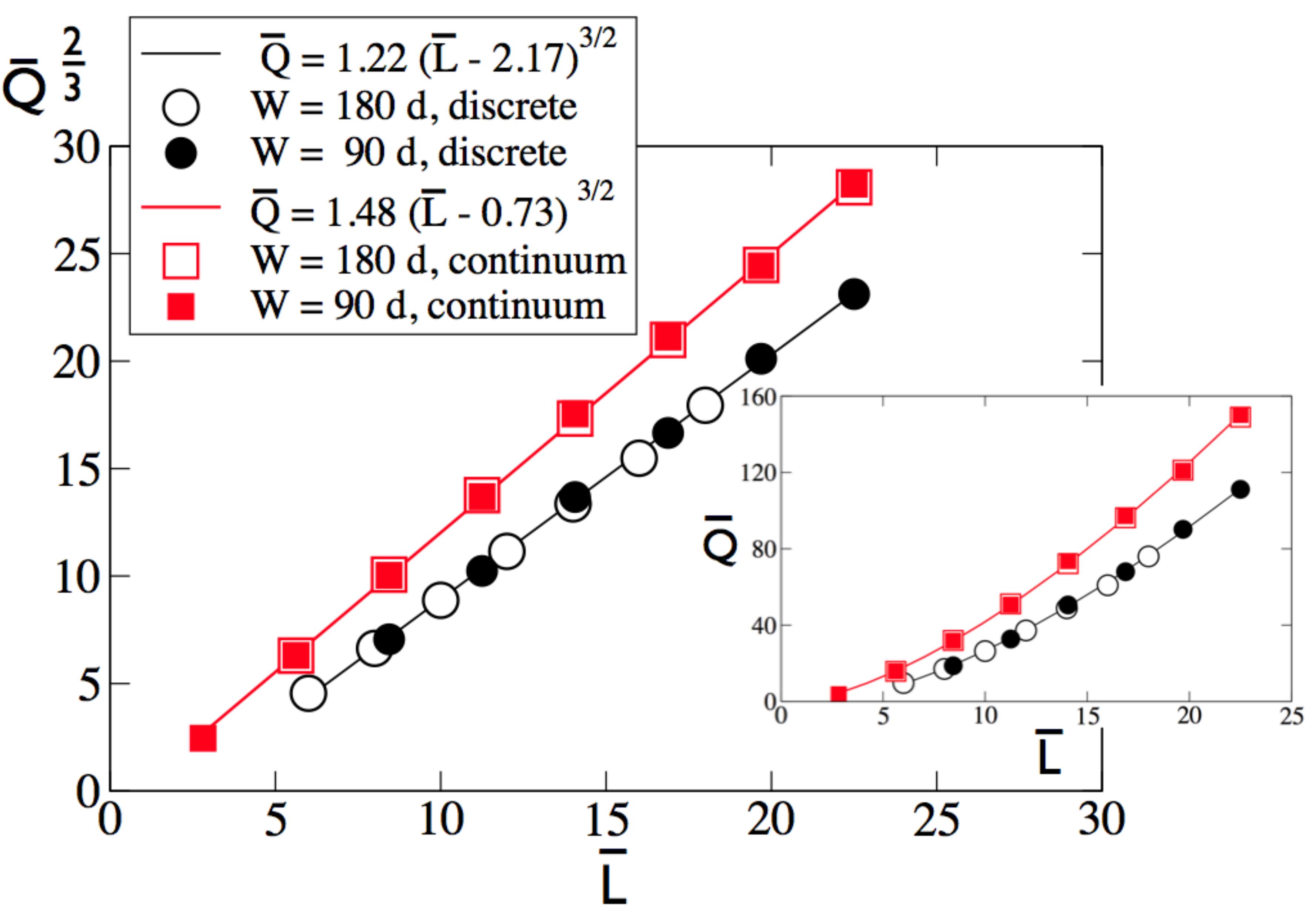}}
\caption{Normalized discharge rate $\bar{Q} = Q/\sqrt{g}d^{3/2}$ as a function of the normalized outlet size $\bar{L}=L/d$ for silos width $W=90d$ and $W=180d$, for discrete and continuum simulations. }
\label{BevSmallLarge}
\end{minipage}
\end{figure*}
%------------------------

\subsection{The visco-plastic silo using a Navier-Stokes solver }
%\subsection{Continuum simulations using the $\mu(I)$-rheology}

The continuum simulations were performed using the Gerris flow solver in two dimensions, which solves  the Navier-Stokes equation for a bi-phasic mixture applying a Volume-Of-Fluid approach \cite{popinet03,popinet09}. The existence of two fluids translates numerically  in different viscosity and density on the simulation grid following the advection of the volume fraction representing the proportion of each fluid. In our case, one fluid stands for granular matter (characterized by the coefficient of internal friction $\mu$) and the other stands for the surrounding air. In \cite{lagree11}, we established that the dynamics of a rapid granular layer is not affected by the surrounding fluid if the latter has a density and viscosity  small enough relatively to those of the granular layer; accordingly, we chose a ratio of $10^{-2}$ between the density and viscosity of the two fluids.  The position of the interface between the two fluids is solved in the course of time based on the spatial distribution of their volume fraction. 
The Navier-Stokes equations:
\begin{eqnarray}
\nonumber
\boldsymbol{\nabla . u} & =& 0 \\\nonumber
\rho \left( \frac{\partial \boldsymbol{u}}{\partial t} + \boldsymbol{u . \nabla u}\right) & =& -\boldsymbol{\nabla} p +\boldsymbol{\nabla}.(2{\boldsymbol{\eta}} \boldsymbol {D})  + \rho \boldsymbol {g} \\ \nonumber
\end{eqnarray}
are thus solved for a two-phase flow, namely granular matter and air, with a variable fraction function $c$:
\begin{eqnarray}
\nonumber
\frac{\partial c}{\partial t} &+& \boldsymbol{\nabla}.(c\boldsymbol{u}) = 0,\\\nonumber
\rho  &=& c\:\rho_\text{grains} + (1-c) \rho_\text{air}, \\\nonumber
\eta  &=& c\:\eta_\text{grains} + (1-c) \eta_\text{air}.\\\nonumber
\end{eqnarray}
As explained in subsection \ref{sec:modA} (relation (\ref{eq:eta})), the viscosity  $\eta_\text{grains} $ of the granular matter is approximated by mean of the friction properties \cite{jop06,chauchat10}: 
\begin{equation}
\eta_\text{grains}  = \min\left(\frac{\mu(I) P}{D_2}, \eta_{max}\right), 
\end{equation}
where $\mu$ is the effective coefficient of friction of the granular flow, $P$ is the local pressure and $D_2$ is  the second invariant of the strain rate tensor $\boldsymbol D$: $D_2=\sqrt{D_{ij}D_{ij}}$. For large values of $D_2$, the viscosity is finite and proportional to $\mu$ and P; when $D_2$ reaches low values, the viscosity $\eta$ diverges. Numerically, this divergence is bounded by a maximum value $\eta_\text{max}$ chosen to be $10^4$ times the minimum value of $\eta$; we have checked that the choice of $\eta_\text{max}$  did not affect the results as long as $\eta_\text{max}$  is large enough (at least $10^2$ times the minimum value of $\eta$). 

\subsection{The discrete silo using Contact Dynamics}

The discrete simulations are performed using the Contact Dynamics algorithm \cite{jean92,radjai09}. The grains are perfectly rigid and obey a strict non-overlap condition at contact. They interact through a Coulombic friction law, imposing that the tangential force at contact $f_t$ is related to the normal force at the same contact $f_n$ through the inequality $|f_t| \leq \mu_c f_n,$  where $\mu_c$ is the coefficient of friction at contact. A coefficient of restitution $e$ prescribes the amount of energy dissipated during collisions. In a given configuration, the algorithm finds all the forces compatible with the constraints,  geometrical and frictional, imposed at each contact. This method has proven a reliable tool to reproduce the behavior of granular matter in many configurations. Further details on the numerical method can be found in \cite{jean92,radjai96,radjai09}.\\
 The simulations discussed in this contribution are performed with a value of the coefficient of restitution set to $e =0.5$, which favor dense flow regimes. The value of the coefficient of friction is set to $\mu_c=0.5$ (glass beads have a contact friction of about $0.2$, and sand grains have coefficient of friction that may vary a lot, but $0.5$ is consistent).  The influence of these two parameters on the flow characteristics was not investigated. Instead, we focus on the silo's geometrical characteristics  to allow comparison with the continuum simulations.

\subsection{Flow configuration}
  The flow configuration investigated, using both continuum or discrete approaches, is a two-dimensional flat-bottomed silo, of width $W$, initial filling height $H$ and  outlet $L$ (see Figure \ref{illus}). The width of the silo is $W=90d$ or $W=180d$; this corresponds to 8066 and 16240 discrete grains respectively.  The initial filling height is $H=90d$, and was no varied. The grains show a slight size dispersity to avoid ordering effect.
In the case of discrete simulations,  the walls of the silo are smooth, and contacts between grains and walls show the same properties ({\it i.e.} same coefficients $e$ and $\mu_c$) as contacts between grains. 
  For the continuum silo, a zero pressure condition is imposed at the top-wall and at the outlet. A no-slip boundary condition is imposed at the side-walls and at the bottom-wall; the effect of this choice compared to free-slip condition is discussed in section~\ref{sec:BC}.\\
  Considering geometrically perfectly identical continuum and discrete granular silos, we can now compare their respective behavior during discharge.
  
%   The grains show a slight size dispersity to avoid ordering effect, too small however to induce segregation: we chose $(d_{max}-d_{min})/d=0.4$, where $d$ is the mean grain diameter ($d=500\mu$m). 

\section{The discharge rate}
\label{sec:dis}

Granular silos have the well-known particularity of releasing their stored material at a constant rate, through a discharge process which, as a consequence, seems independent of the mean pressure inside the silo itself.  This particularity is captured by the Berverloo scaling~\cite{beverloo61}, which relates the discharge rate $Q$ to the silo outlet $L$ according to the following scaling, provided $L$ is large enough\cite{zuriguel05,mankoc07}:
\begin{equation} 
Q = C \sqrt{g} (L-kd)^{N-1/2},
\label{eq:beverloo}
\end{equation}
where $N$ is the dimension of the problem, and $C$ and $k$ are non-dimensional constants.  The constant $k$ represents the volume of exclusion due to steric constraints applied by the rigid grains,  reducing the effective size of the outlet by a multiple of the grain diameter $d$; a typical value for $k$ is 2.
The constant $C$ is typically of $0.5$ in 3D \cite{beverloo61}.
Although the range of validity of the Beverloo scaling is bounded, small and very large apertures inducing different behaviors \cite{mankoc07}, it is surprisingly robust, and was recovered both numerically and experimentally for a great variety of granular matter \cite{potapov96,choi05,bartos06,sheldon10,aguirre11,gonzalez11,hilton11}. \\
The physical origin of this scaling is often attributed to the Janssen effect, namely a pressure screening of the lower region of the silo due to the mobilization of friction forces at the walls \cite{janssen95,sperl06,walker66,ovarlez03,perge12}.  In contradiction to this  explanation, experimental works have shown that the Beverloo scaling holds in configurations where Janssen effect could not be active \cite{sheldon10,aguirre11}.  Recently, continuum simulations of the silo discharge using the continuum approach  applied in this paper suggest that the Beverloo scaling results from the yield stress properties of the material \cite{staron12}, in agreement with \cite{nedderman92}.
 This aspect will not be discussed any further here.  Instead, we compare the discharge of the discrete granular silos simulated by contacts dynamics and the discharge of continuum granular silos simulated by the solver Gerris with the $\mu(I)$-flow-law, and focus on the consistency between the two approaches.

\subsection{Recovering the Beverloo scaling}

In a first set of simulations, we perform series of silo discharges with $W=90d$, and  with outlet size $L$ exactly similar in the discrete and the continuum cases, in order to allow direct comparison.   We consider  $L$ varying from $L=5.63d$ to  $L= 22.5d$.   The evolution of the volume $V$ left in the silo (normalized by the initial volume $V_0$) is reported in Figure \ref{Discharge} as a function of time (normalized by $\sqrt{d/g}$). For both discrete and continuum cases, the discharge rate is constant, in agreement with experimental observation.
 The normalized discharge rate $\bar{Q} = Q/\sqrt{g}d^{3/2}$ can thus be plotted as a function of  $\bar{L} = L/d$.  For both discrete and continuum silos, we recover the Berverloo scaling (Figure \ref{Discharge}, inset):
\begin{eqnarray}
\label{bevdis}
{Q} &=& 1.22 \sqrt{g} (L-2.17d)^{3/2} \text{ for discrete},\\
\label{bevcon}
{Q} &=& 1.55 \sqrt{g} (L-0.56d)^{3/2} \text{ for continuum}. \nonumber
\end{eqnarray}
The Beverloo scaling was recovered in many instances using discrete simulations \cite{langston95,potapov96,hirshfeld97,bartos06,gonzalez11,djouwe12}, showing  the robustness of this flow behavior. Observing this typical granular phenomenology in the case of the discharge of a viscous flow, though non-Newtonian, is however not trivial \cite{staron12}. \\%
No prior adjustment of the rheological parameters  was made: we use  $\mu_s = 0.32$,  $\Delta \mu=0.28$ and $I_0=0.4$ for the continuum simulations, $\mu_c=0.5$ and $e=0.5$ for the discrete simulations.  Hence we do not expect the two approaches to coincide quantitatively at this stage. Nevertheless, beside the fact that the continuum discharge is more rapid for a given $L$ (corresponding to a large pre-factor $C$), two important differences can be noted.
Expectedly,  the grains rigidity in the discrete case leads  to a lower effective outlet $(L-kd)$: $k=2.17$ for discrete simulations, while $k=0.56$ for continuum (see scalings (\ref{bevdis})  and (\ref{bevcon})). Moreover, for a given outlet size $L$, the discrete discharge is more efficient: $\simeq 95\%$ of material evacuated, against $\simeq 80\%$ for the continuum discharge.

%---------VELOCITY--------------------

 \begin{figure}
\begin{minipage}{1.\linewidth} 
\centerline{\includegraphics[angle = -90,width = 1.\linewidth]{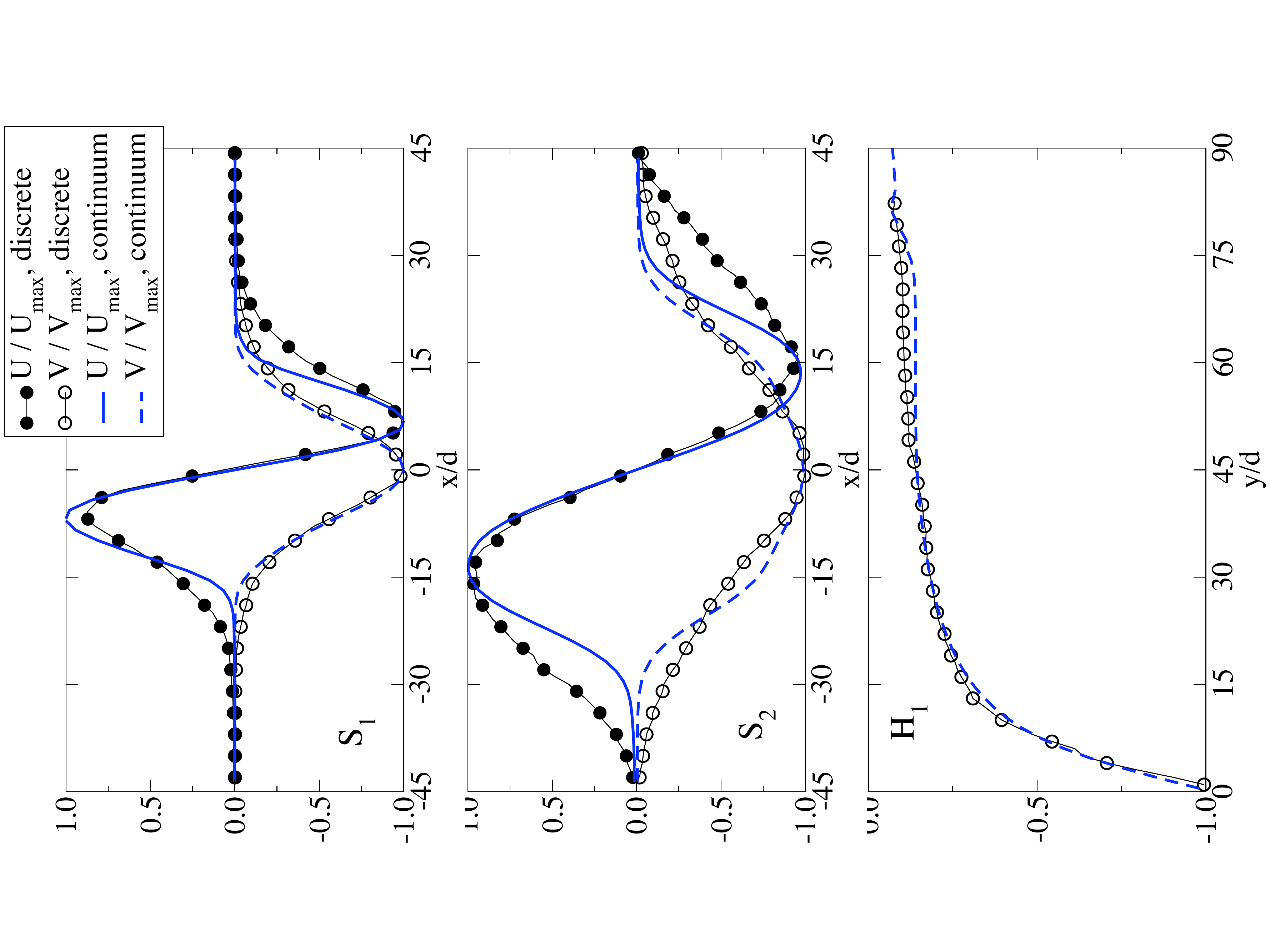}}
\end{minipage}
\caption{Horizontal and vertical velocity profiles for granular and continuum silos along cross-sections $S_1$ and $S_2$, and vertical axis $H_1$, rescaled by the velocities maximum values. (See Figure \ref{illus} for localization of sections  $S_1$ and $S_2$ and $H_1$).}
\label{VelSH}
\end{figure}

 \begin{figure*}
\begin{minipage}{0.32\linewidth} 
\centerline{\includegraphics[angle = -90,width = 1\linewidth]{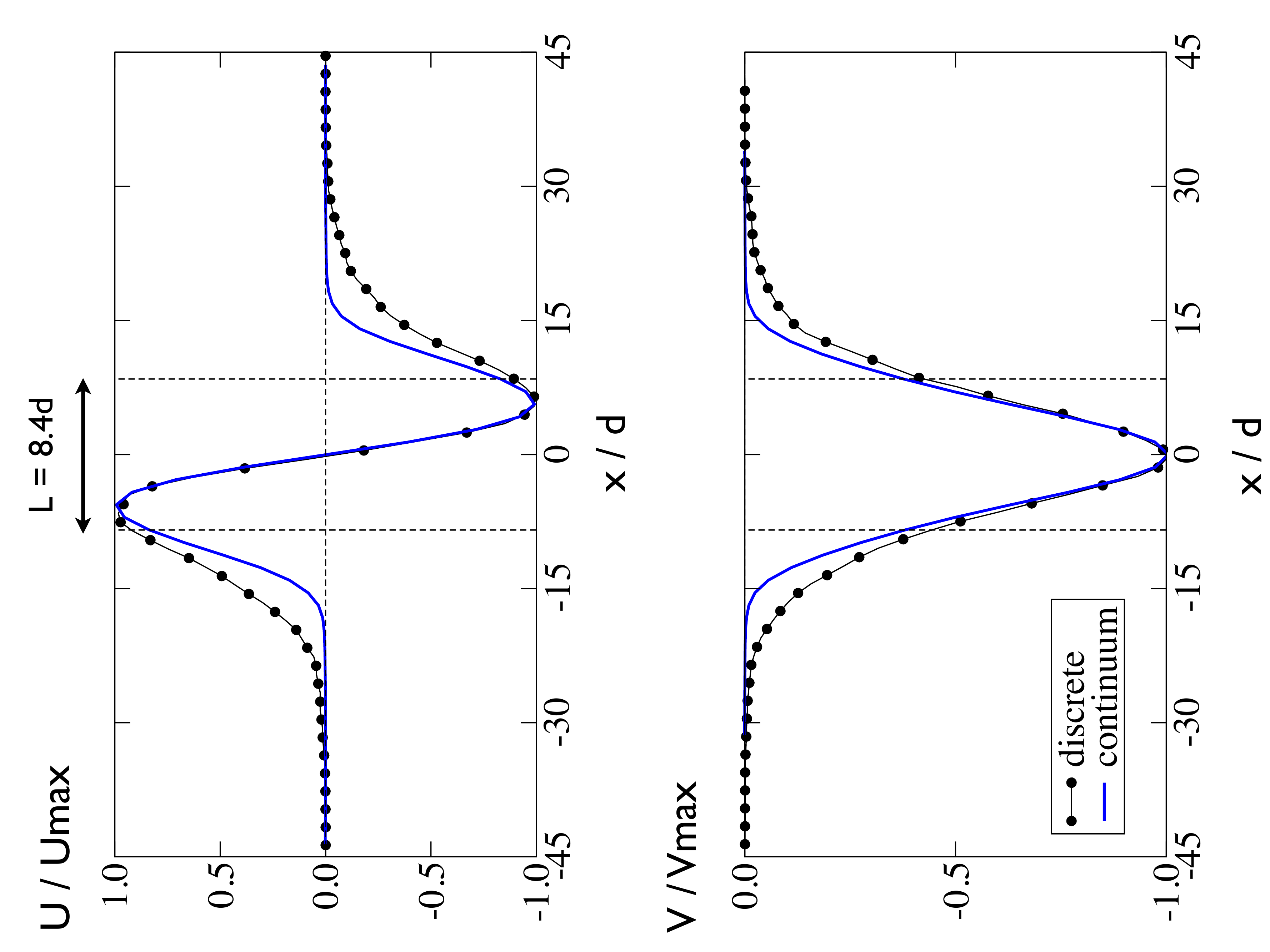}}
\end{minipage}
\hfill
\begin{minipage}{0.32\linewidth} 
\centerline{\includegraphics[angle = -90,width = 1\linewidth]{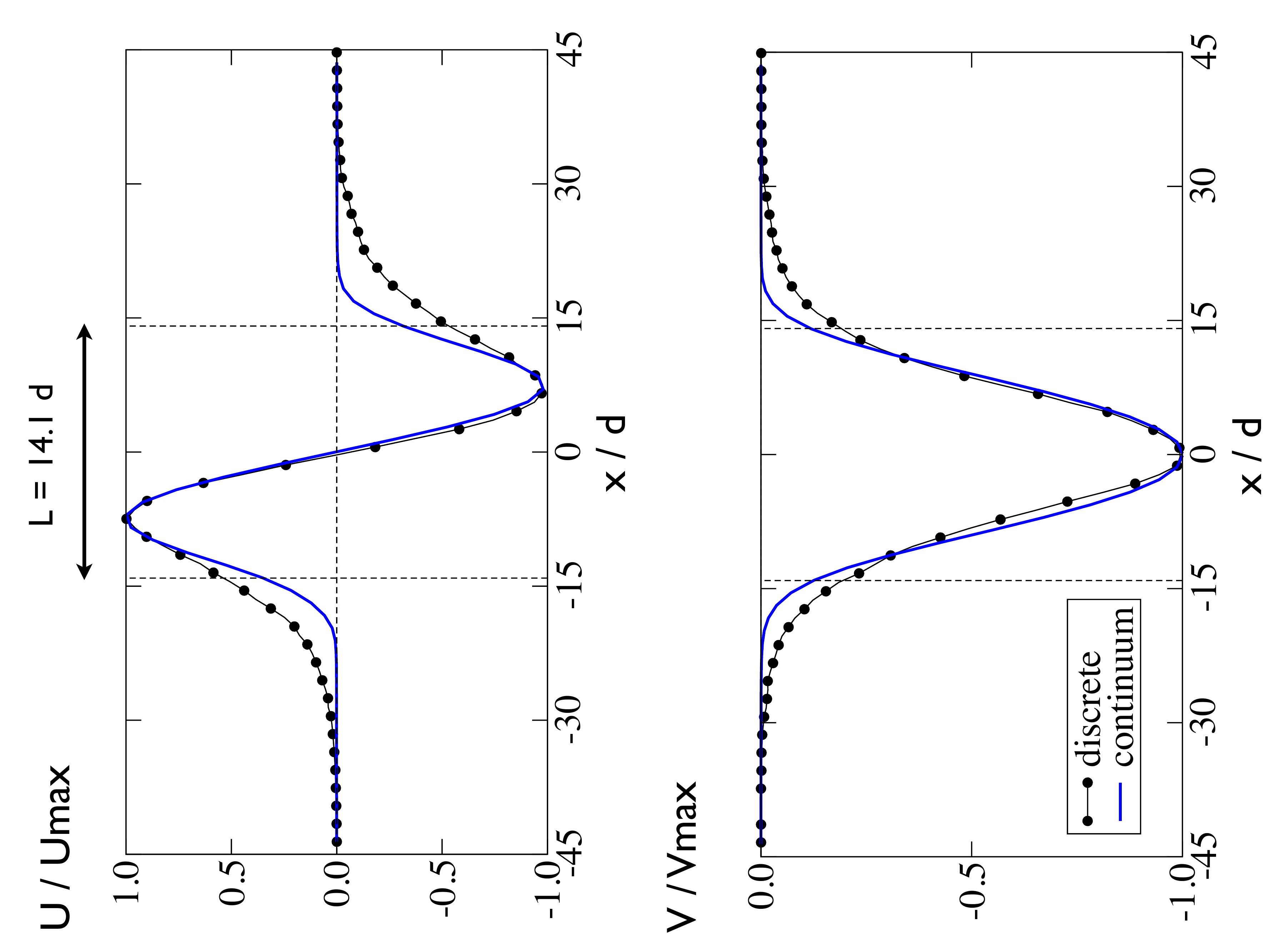}}
\end{minipage}
\hfill
\begin{minipage}{0.32\linewidth} 
\centerline{\includegraphics[angle = -90,width = 1\linewidth]{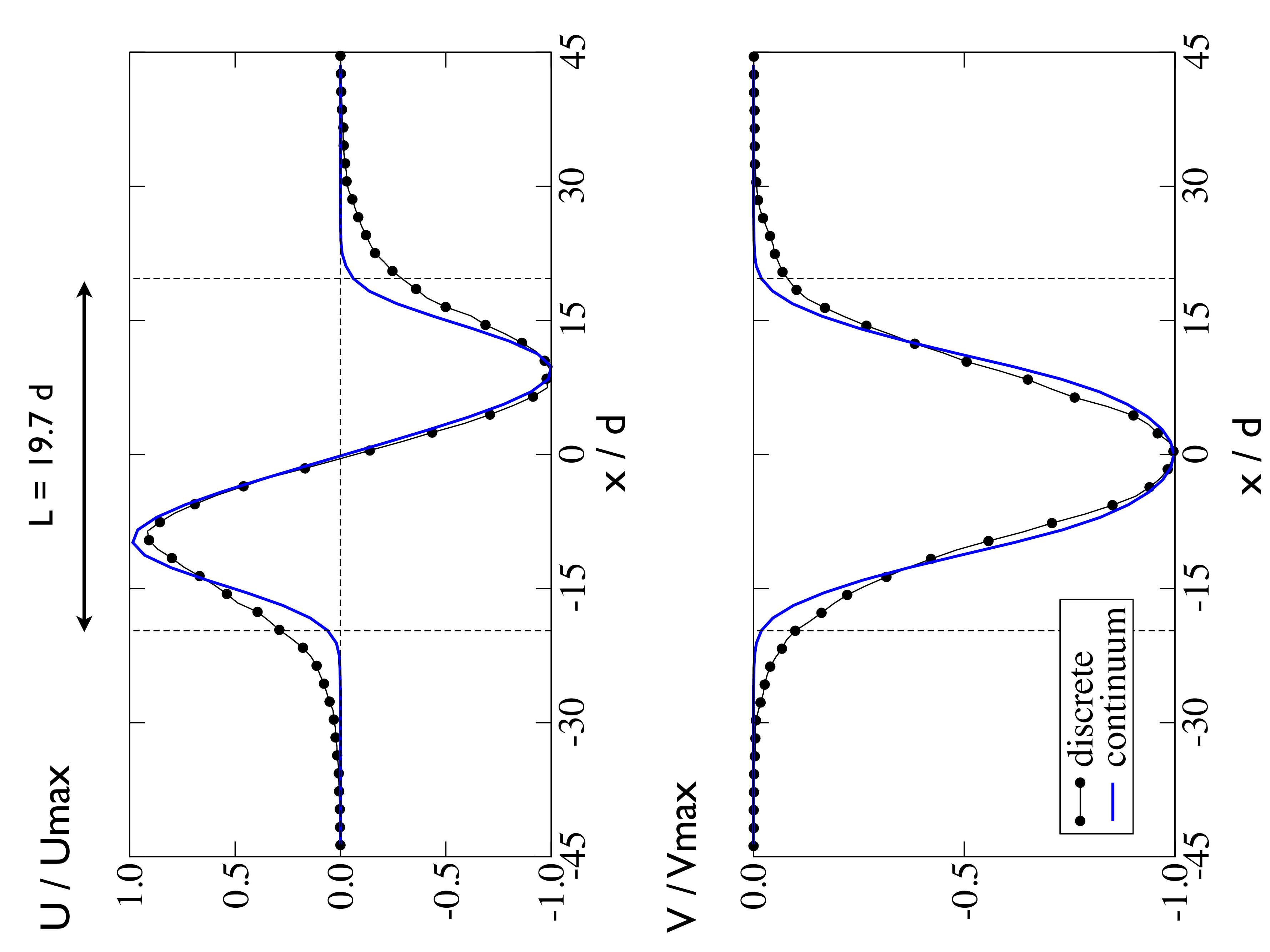}}
\end{minipage}
\caption{Horizontal ($U$) and vertical ($V$) velocity profiles along  the cross-section $S_1$, normalized by their maximum values $U_\text{max}$ and $V_\text{max}$ respectively,  for granular and continuum silos with outlet size $L=8.4 d$ (left),   $L=14.1 d$ (middle) and $L=19.7 d$ (right).}
\label{UV}
\end{figure*}
\begin{figure}
\begin{minipage}{0.99\linewidth} 
\centerline{\includegraphics[width = 1.\linewidth]{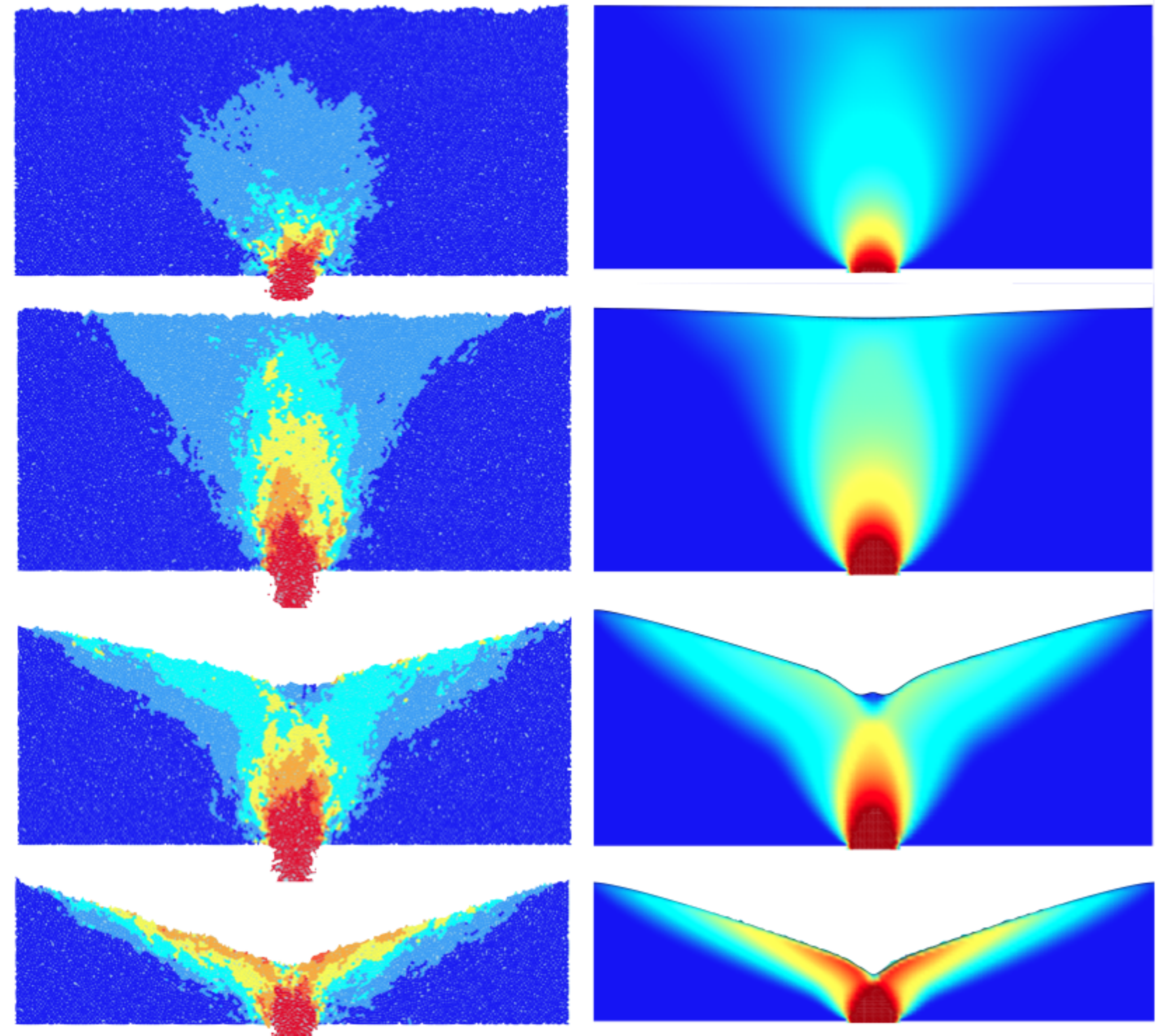}}
\end{minipage}
\caption{(Color On-line) Velocity field for discrete simulation and continuum simulation with $L=16.8d$ throughout the discharge process. The color scale is linear, with an upper-bound value shown in red color.}
\label{VelMap}
\end{figure}
 %------------------------------------

\subsection{Influence of the silo's width}

To check the influence of the silo's width $W$ (mobilization of friction forces at the walls being often resorted to as explanation for the Beverloo scaling), we perform series of simulations with larger silos with $W=180d$ instead of $W=90d$.  As previously, for both discrete and continuum simulations, we plot the normalized discharge rate $\bar{Q}$ as a function of the normalized outlet size $\bar{L}$ in Figure \ref{BevSmallLarge} for $W=180d$, together with the results obtained for $W=90d$. We observe that the value of $W$, at least in the range considered here, has no effect on the discharge rate.  This result tends to show that the discharge is dominated by local factors, rather than by the state of the system at the walls. \\
The influence of the initial height of material stored in the silo was not investigated here. While we expect the latter to have no influence on the silo discharge in the case of discrete simulations, this is not as obvious in the case of continuum simulations. This aspect was studied in detail in \cite{staron12}, showing that the influence of initial height  in the continuum silo discharge is weak.

%--------------PRESSURE-------------

 \begin{figure*}
\begin{minipage}{0.323\linewidth} 
\centerline{\includegraphics[angle = 0,width = 1.05\linewidth]{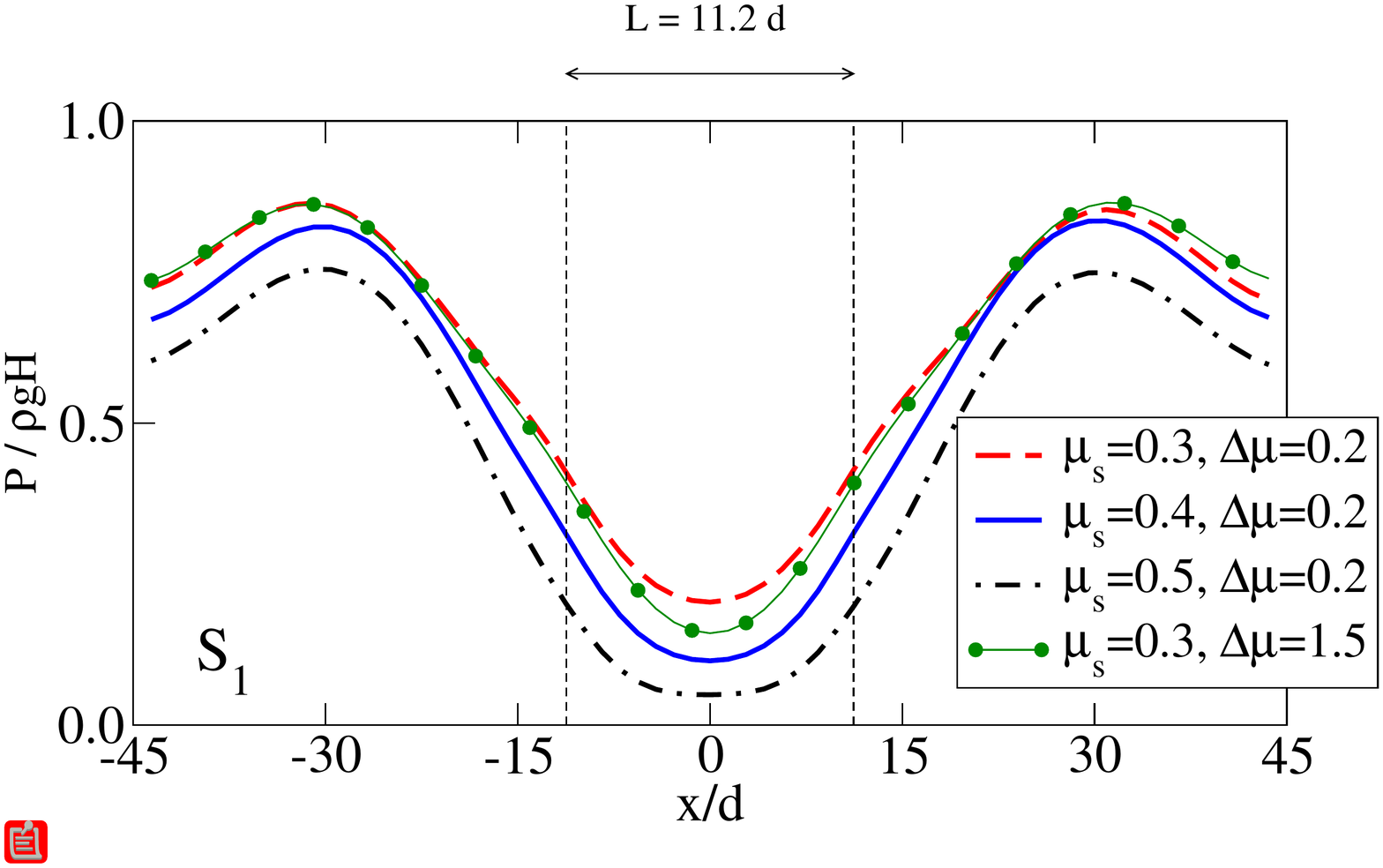}}
\end{minipage}
\hfill
\begin{minipage}{0.323\linewidth} 
\centerline{\includegraphics[angle = 0,width = 1.05\linewidth]{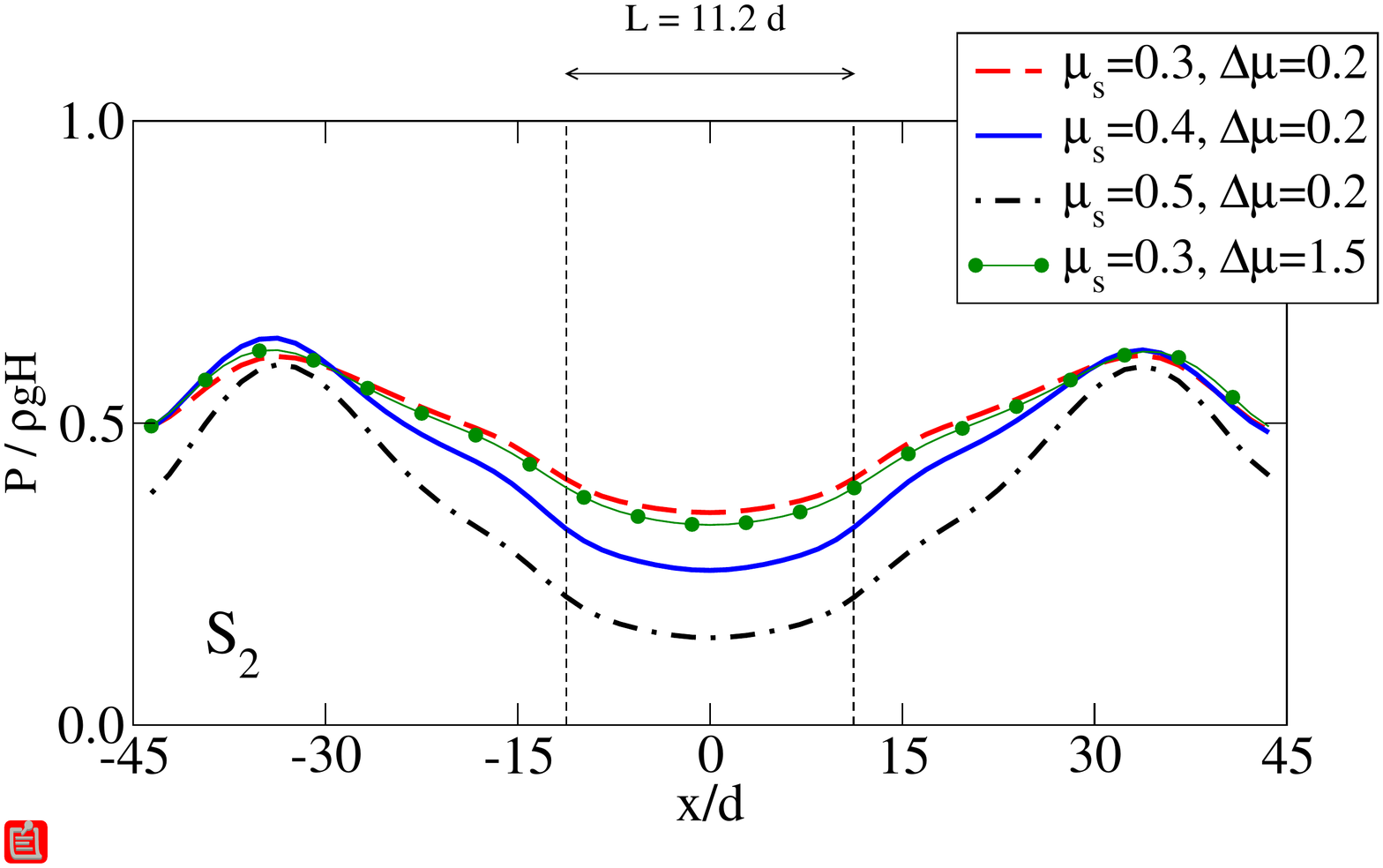}}
\end{minipage}
\hfill
\begin{minipage}{0.323\linewidth} 
\centerline{\includegraphics[angle = 0,width = 1.05\linewidth]{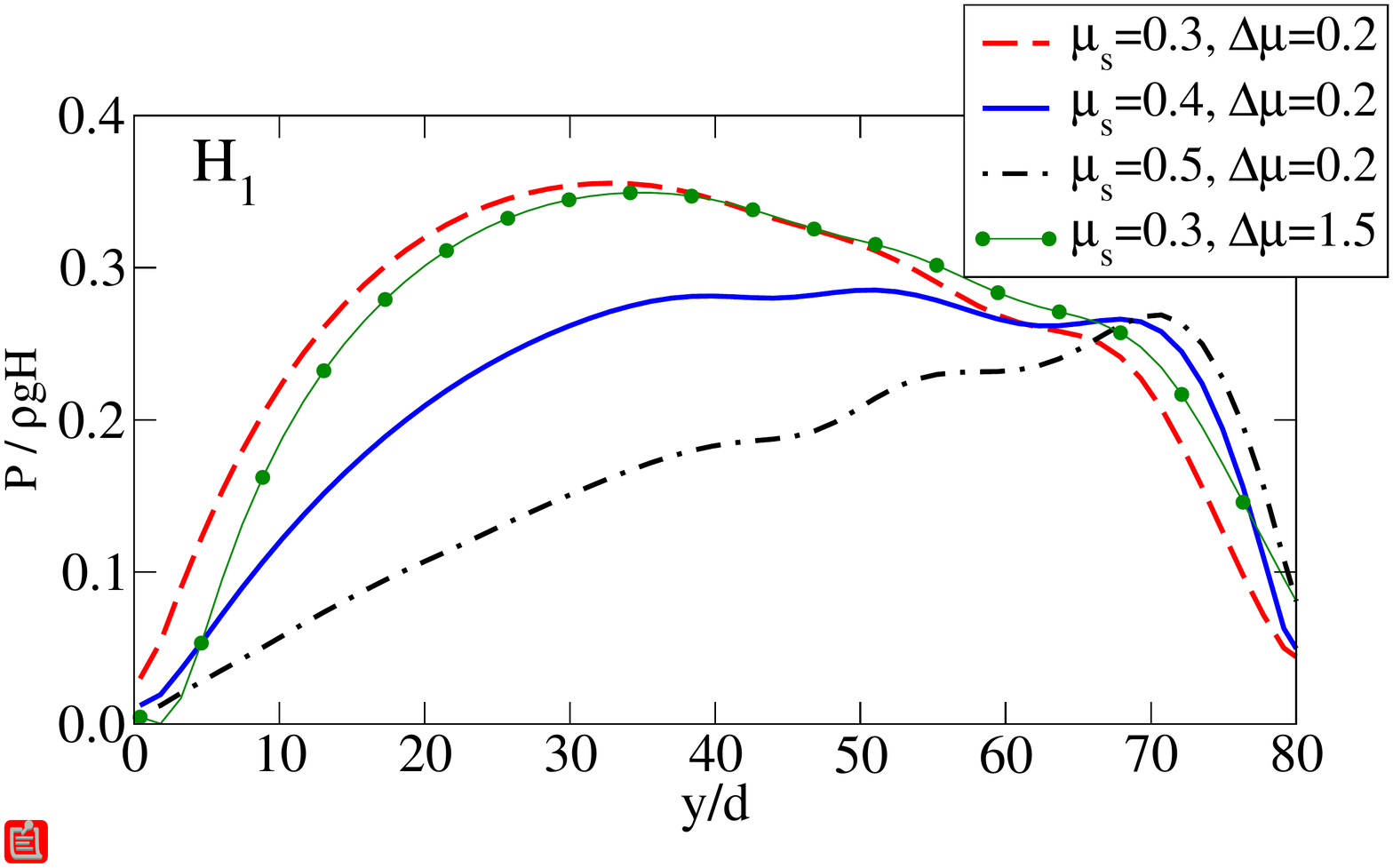}}
\end{minipage}
\caption{To write: Horizontal ($U$) and vertical ($V$) velocity profiles along  the cross-section $S_1$, normalized by their maximum values $U_\text{max}$ and $V_\text{max}$ respectively,  for granular and continuum silos with outlet size $L=8.4 d$ (left),   $L=14.1 d$ (middle) and $L=19.7 d$ (right).}
\label{Pmu}
\end{figure*}

\begin{figure}
\begin{minipage}{1.\linewidth} 
\centerline{\includegraphics[angle = -90,width = 1.\linewidth]{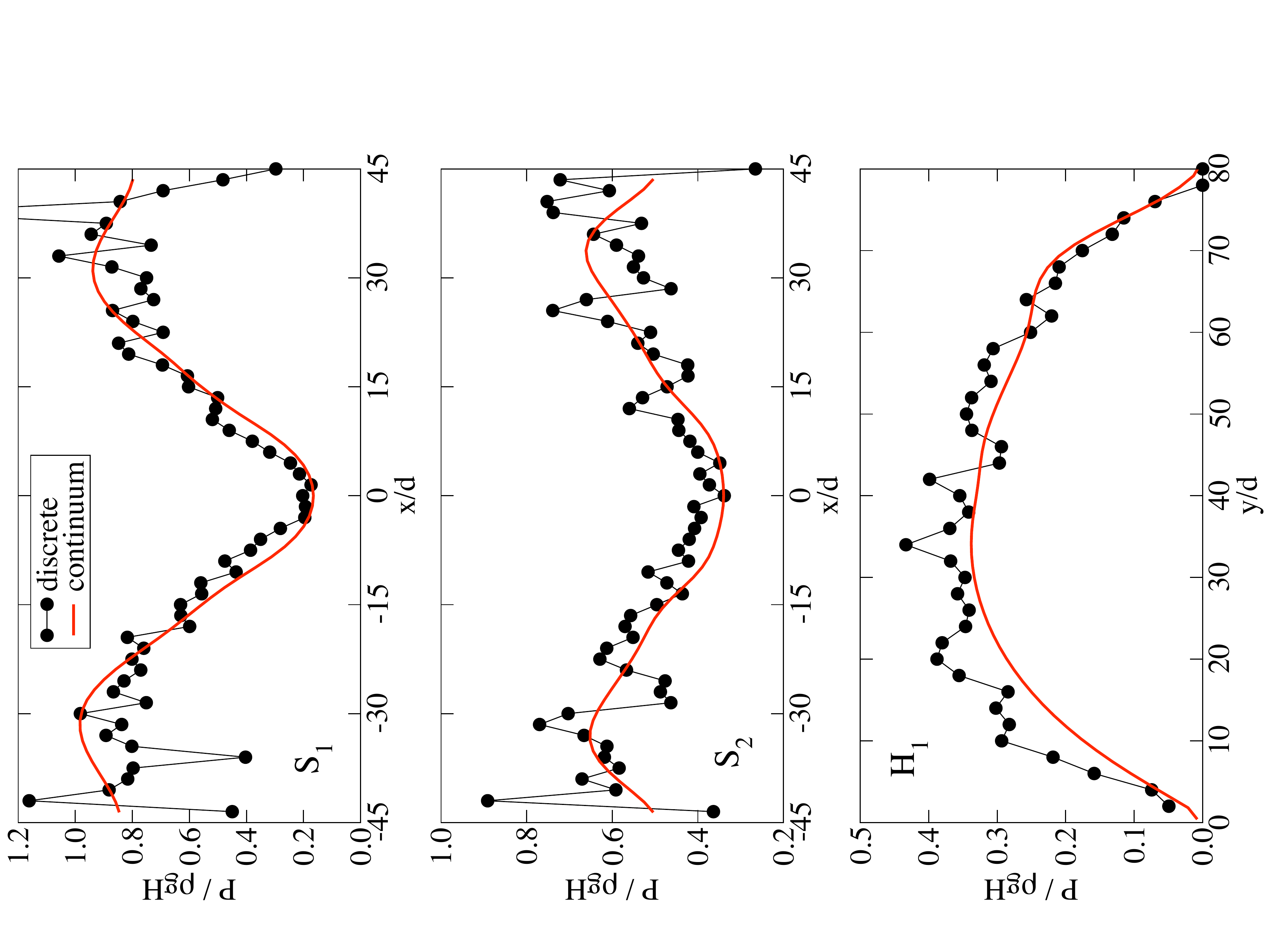}}
\end{minipage}
\caption{ Profile of the pressure $P$ (normalized by $\rho g H$) along the horizontal  cross-sections $S_1$ and $S_2$ and vertical axis $H_1$ (at time $T= 10 \sqrt{gd}$), for the continuum (plain line) and the discrete  ($\bullet$ symbols)  simulations of granular silos with $L=11.2d$. (See Figure \ref{illus} for localization of $S_1$, $S_2$ and $H_1$).}
\label{PSH}
\end{figure}

\begin{figure}
\begin{minipage}{0.99\linewidth} 
\centerline{\includegraphics[width = 1.\linewidth]{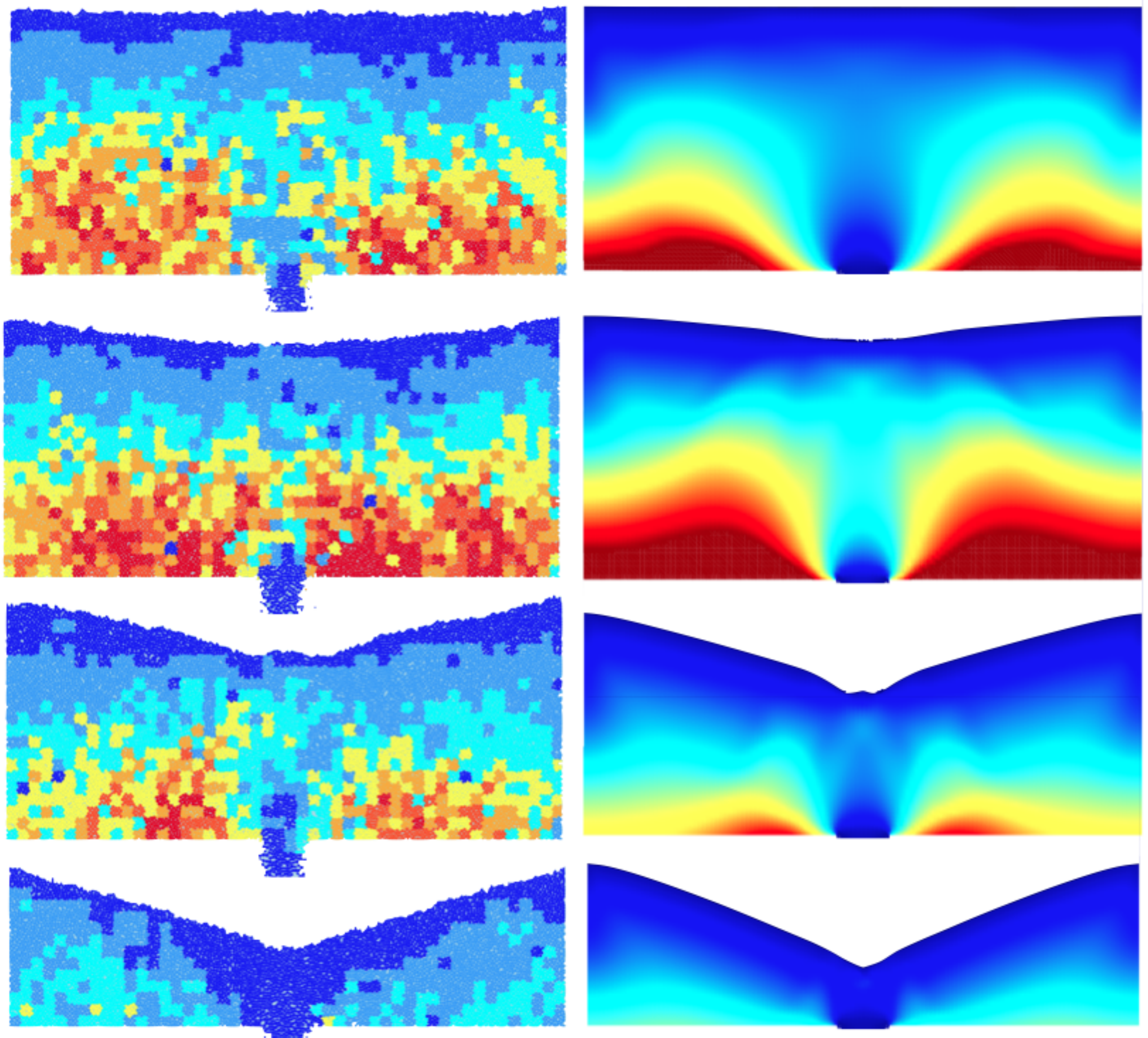}}
\end{minipage}
\caption{(Color On-line) Pressure field for discrete and continuum simulations with $L=16.8d$ throughout the discharge process. The color scale is linear, with an upper-bound value of $0.78 \rho gH$ (in red, color on-line). }
\label{PMap}
\end{figure}
%--------------------------

% -----------------------------------------
\section{The velocity field}
\label{compvel}

From the shape of the Beverloo scaling, and according to intuition, it is clear that the discharge velocity and the velocity field in the bulk of the silo will depend on the value of the rheological parameters adopted. Since no prior adjustment was performed to make the continuum and discrete discharge coinciding quantitatively (this non-trivial aspect being discussed in details in section \ref{tunemui}), quantitative comparison of the velocity field for the two approaches is not possible at this stage.  However, qualitative comparison of the shape of the velocity field for different apertures and along different profiles is possible. 
Figure \ref{VelSH} shows the horizontal and vertical velocities profiles along the cross sections S$_1$ and S$_2$, as well as along the vertical axis H$_1$ (see Figure \ref{illus} for locating S$_1$, S$_2$ and H$_1$), for an outlet size $L=11.2d$, at $T=10\sqrt{d/g}$ and after normalization of the horizontal and vertical velocities by their maximum values  $U_\text{max}$ and $V_\text{max}$ respectively.  The shape of the horizontal profiles S$_1$ and S$_2$ reproduce the shape already observed elsewhere \cite{choi05,rycroft09}. 
Continuum and discrete models show a reasonable qualitative agreement right above the outlet (profiles along $S_1$), but tend to differ in the area of the outlet edges, where the velocity decreases towards zero, and higher in the bulk (profiles along $S_2$). The transition from a flowing state to a static one is sharper in the continuum systems, while discrete systems exhibit a larger area of slow shear before freezing in a static state. Yet, the area of rapid flow is well reproduced by the  continuum model.
Figure~\ref{UV} shows the velocity profiles along the cross section S$_1$ for the three cases $L=8.4 d$, $L=14.1 d$ and $L=19.7 d$. The same conclusions apply:  the shape of the rapid flow is well captured by the continuum model, but areas of slower motion tend to differ.\\ % Note that in this effect may be partly accounted for by the rigidity of the discrete grains: indeed, the effective size of the silo outlet is smaller for discrete grains than for "continuous" ones, thus creating a  \\
For visual inspection, Figure \ref{VelMap} shows snapshots of the velocity field during the discharge of a discrete and a continuum silo with $W=180d$ and $L=16.9d$, at different instants, and using the same color scale. \\

%--------- SURFACE AND DEFORMATIONS-----------------------
\begin{figure}
\begin{minipage}{1.\linewidth} 
\centerline{\includegraphics[angle = 0,width = \linewidth]{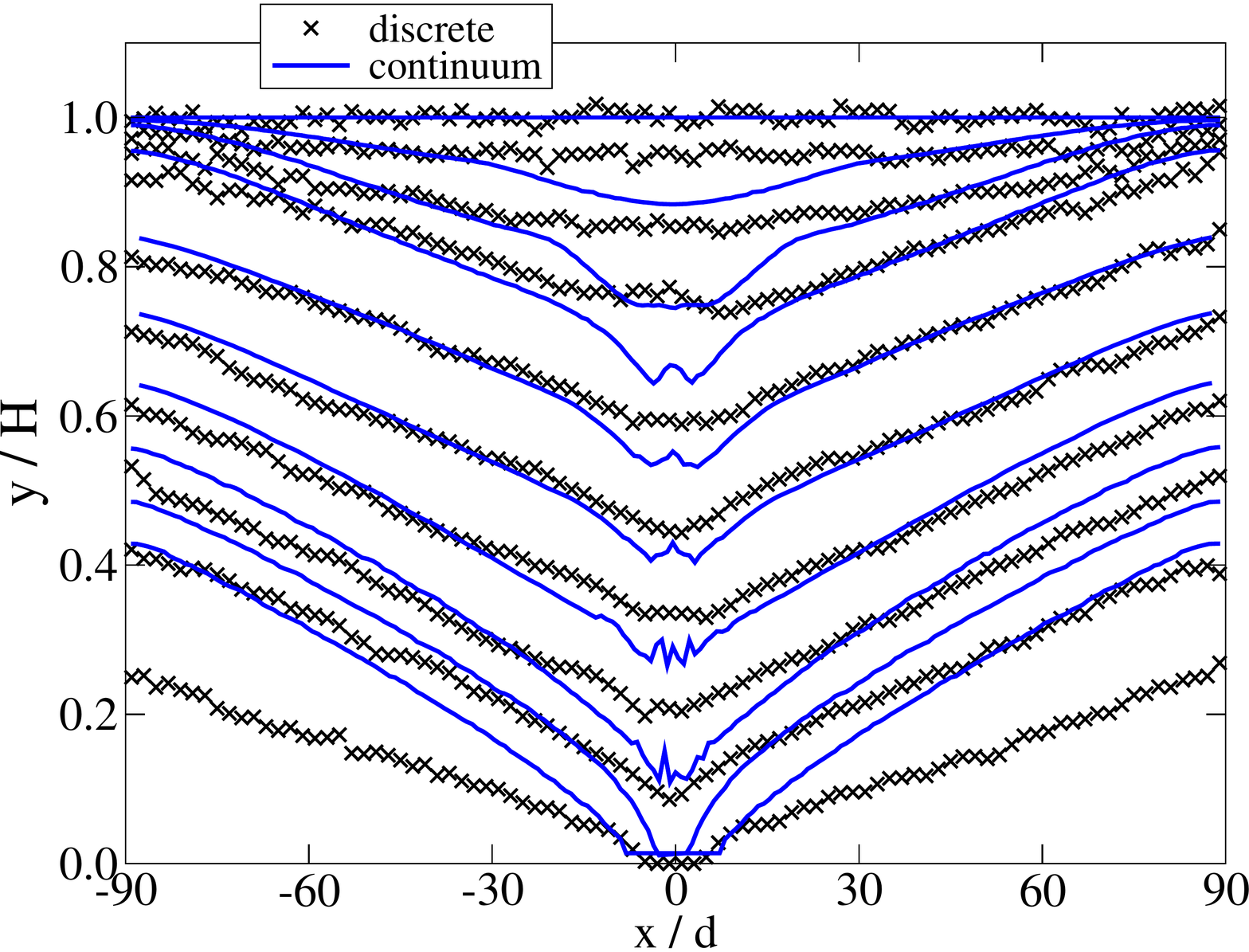}}
\end{minipage}
\caption{Shape of the free surface during the discharge of identical discrete and continuum silos at successive instants of the discharge process $t/T_0 = 0$, $0.05$, $0.1$, $0.15$, $0.25$, $0.35$, $0.45$, $0.55$, $0.65$  and in the final state, where $T_0$ is the total duration  of the discharge. The outlet is the same in both cases: $L = 16.8d$. }
\label{surf}
\end{figure}

\begin{figure}
\begin{minipage}{1.\linewidth} 
\centerline{\includegraphics[angle = -90,width = 1.\linewidth]{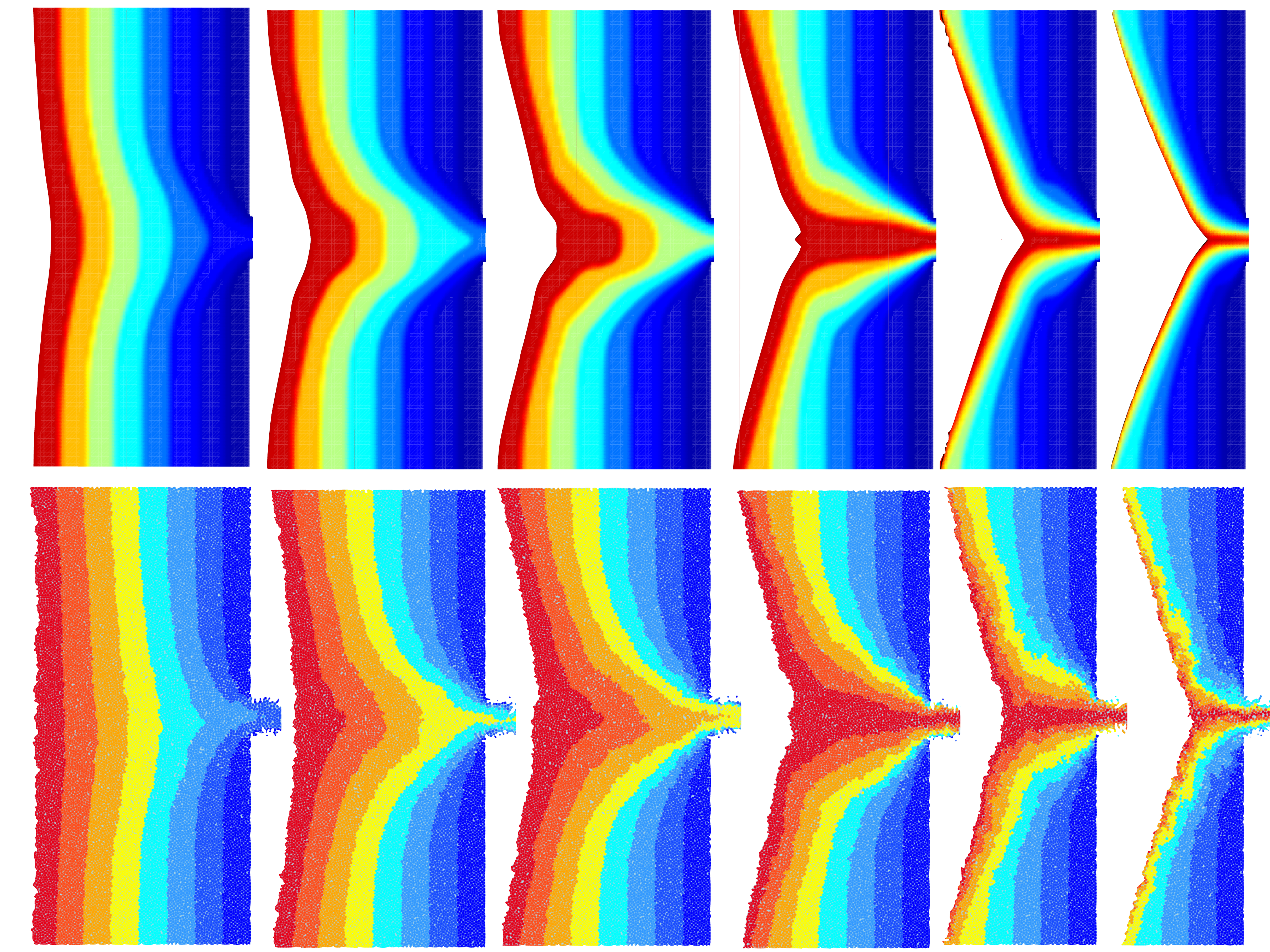}}
\end{minipage}
\caption{(Color-Online) Inner deformations in granular discrete (left) and continuum (right) silos with W=180d, and $L= 16.9d$, for $\bar{t} = t/T_{0}= 0.06, 0.12, 0.17, 0.26, 0.56$ and $0.62$, where $T_0$ is the duration  of the discharge. The  different colors are used as tracers.}
\label{stripes}
\end{figure}
%----------------------------------------------------------------------------------

%------------------------------------------
\section{The pressure field}
\label{comppre}

The pressure field in a silo is known to obey a non-trivial distribution due to the presence of confining walls and the existence of a low pressure condition created by the outlet \cite{janssen95,perge12}. Another factor affecting the pressure field are the yield stress properties of the material \cite{staron12}. This is illustrated in Figure~\ref{Pmu} where the pressure profiles along the cross sections S$_1$ and S$_2$ and along the horizontal axis H$_1$ are plotted for a continuum discharge using different values of the rheological parameters $\mu_s$ and $\Delta \mu$. Independently of the values chosen, the pressure field exhibits strong variations at a given height, with a marked minimum above the outlet even at a significant distance away from it (as along section S$_2$). Increasing the static frictional properties ($\mu_s$) decreases the pressure in the bulk and close to the outlet, thereby decreasing local pressure gradient, hence discharge velocity. Increasing the dependence on the inertial number (namely increasing $\Delta \mu$ for a given $\mu_s$) allows to decrease the pressure only in the areas of higher shear, namely in the area of the outlet. As will be seen in section \ref{tunemui}, this difference is of importance in the perspective of adjusting parameters to achieve quantitative agreement between continuum and discrete granular systems. \\
As a consequence,  quantitative comparison of the pressure field in discrete and continuum silos is not readily possible; however, a qualitative comparison of pressure profiles is shown in Figure \ref{PSH} for silos of outlet $L=11.2d$, along the cross-sections S$_1$, S$_2$ and vertical axis H$_1$, at $T=10\sqrt{d/g}$ for the rheological parameters considered so far  (i.e.  $\mu_s = 0.32$,  $\Delta \mu=0.28$ and $I_0=0.4$ for the continuum simulations, $\mu_c=0.5$ and $e=0.5$ for the discrete simulations).  The result for discrete simulations is averaged over a larger time-window in order to reduce (but not supress) the large fluctuations characteristic for granular matter.  These fluctuations are much higher in the static zone (i.e. closer to the walls), where force chains can form through enduring contacts; areas of rapid flow (i.e. closer to the outlet) where contacts are short-lived, are much smoother. The agreement appears fairly good. Discrete and continuum simulations share the following features: a marked dip of pressure above the outlet,  the existence of two high pressure regions on one and the other side of the outlet, and a slighter decrease close to the walls.  \\
Maps of the pressure field are shown  in Figure \ref{PMap} at different instants of the discharge for both discrete and continuum systems. In the discrete case, the pressure is averaged over squares of 7 grains diameters sides. 
%

%-------------------------------------------
\section{Inner and free-surface deformations}
\label{sec:def}

To compare the geometry of the discrete and continuum systems in the course of their discharge, we focus on the shape of the surface and on the inner-deformation at equivalent moments, that is at equal fractions of the discharge duration $T_0$. To allow for better sampling, we consider wide configurations: $W=180d$. The outlet size is $L=16.9d$ for both discrete and continuum simulations\\
 Figure \ref{surf} shows the geometry of the free surface for  continuum and discrete granular silos at  $t/T_0 = 0$, $0.05$, $0.1$, $0.15$, $0.25$, $0.35$, $0.45$, $0.55$, $0.65$  and in the final state.   The general trend reflects the fact that the discharge occurs at constant rate in both cases. Noticeable differences emerge early in the flow:  while the surface of the discrete system remains flat  in the first moments and starts forming well-defined slopes only  at a later stage, the surface of the continuum system soon evolves into a dip. The latter however eventually vanishes.  The agreement  increases while the discharge proceeds, with the difference that continuum slopes are slightly convex in shape when discrete slopes remain straight. An important difference arises in the final state:  more material remains in the continuum silo than in the discrete one (as already observed from Figure \ref{Discharge}). Indeed, by successive avalanches, discrete grains are more efficient at flowing out of the container. As a consequence, the slopes of the remaining material in the continuum case are much steeper than in the discrete case.
One may suspect the wall conditions (smooth frictional for discrete simulations, no-slip for continuum) to play an important role in this aspect; however, as will be seen in section \ref{sec:BC}, changing wall-conditions only marginally affects these features.\\
 Figure \ref{stripes} shows the inner deformations occurring during the discharge at different moments using colors to trace down the particles (either grains or fluid volumes). The general agreement between the two is good. Note that the singularity observed above the outlet in the continuum case for $t/T_0=0.26$ was observed experimentally in \cite{samadani02}, and is also visible in discrete simulations in the shape of a little mounded swell (also visible in Figure \ref{PMap}).

%------ TUNE MU I
 
\begin{figure}
\begin{minipage}{1.\linewidth} 
\centerline{\includegraphics[angle = 0,width = 0.9\linewidth]{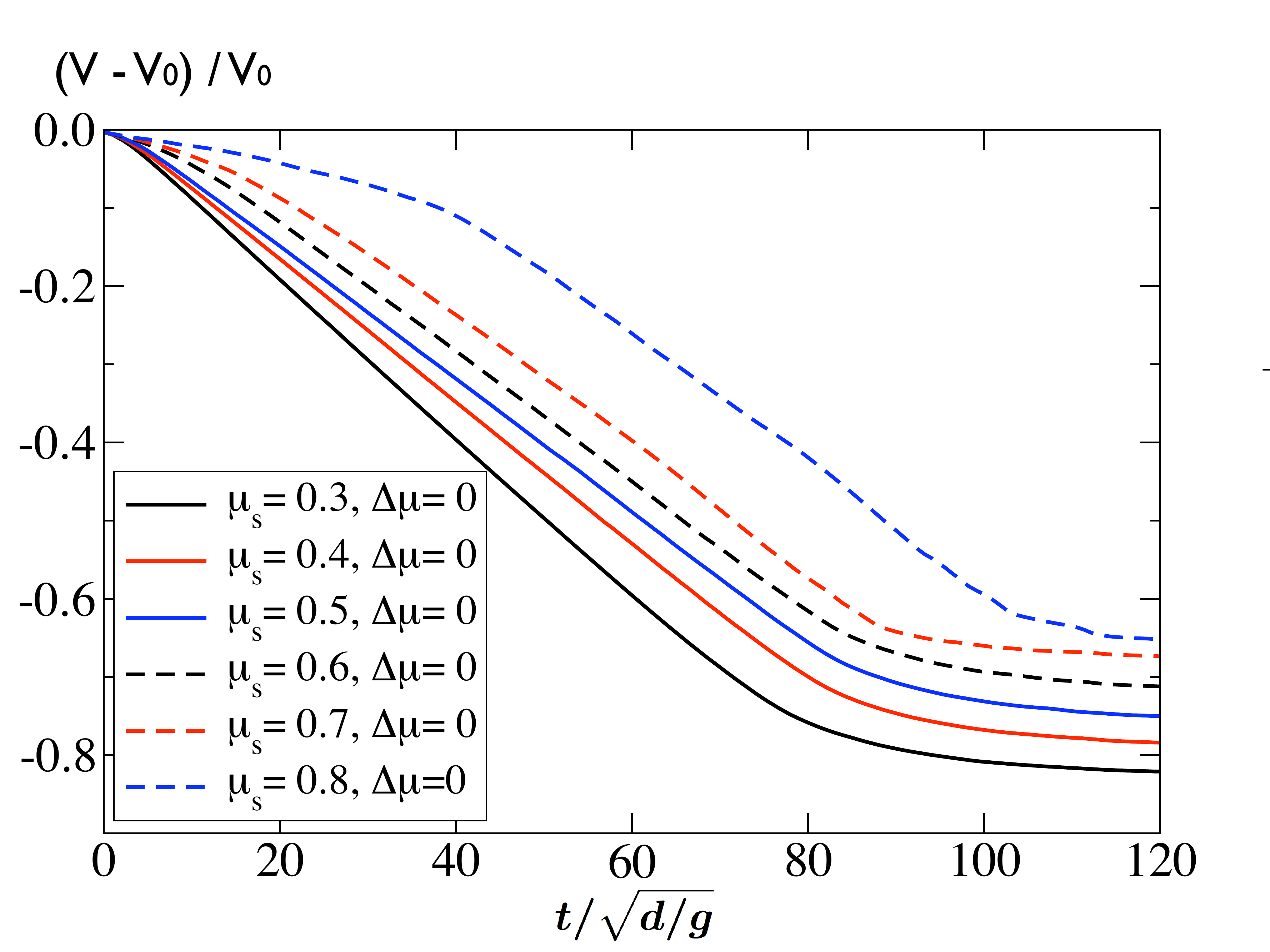}}
\end{minipage}
\caption{(Color on-line) Normalized volume of  matter left in the silo as a function of the normalized time $t/\sqrt{d/g}$ for  continuum simulations with different values of the static coefficient of friction $\mu_s=0.3$, $0.4$, $0.5$, $0.6$, $0.7$ and $0.8$, with $\Delta \mu = 0.$ ($L=14.1d$). }
\label{TuneDis1}
\end{figure}
\begin{figure}
\begin{minipage}{1.\linewidth} 
\centerline{\includegraphics[angle = - 0,width = 0.9\linewidth]{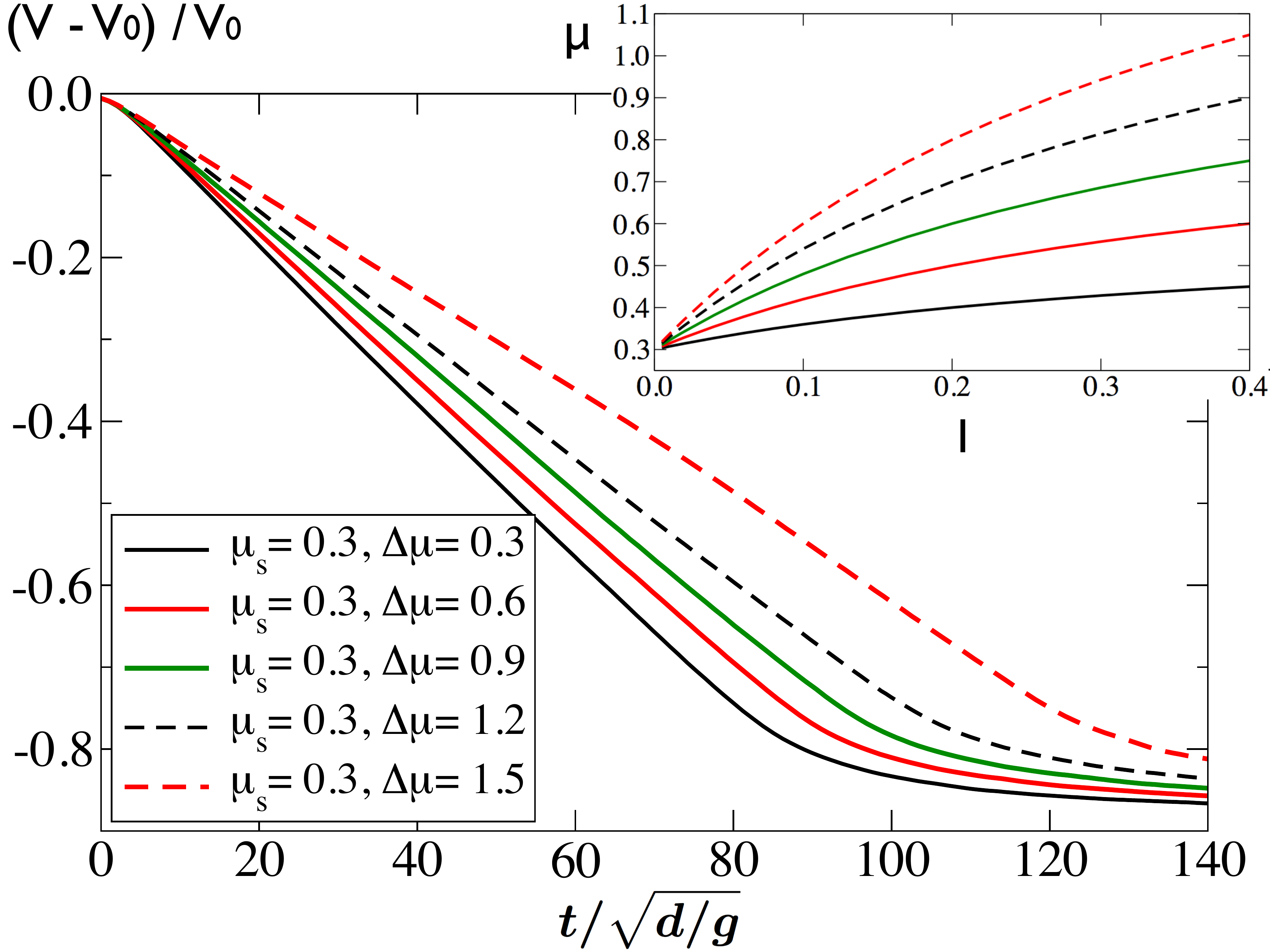}}
\end{minipage}
\caption{(Color on-line) Normalized volume of  matter left in the silo as a function of the normalized time $t/\sqrt{d/g}$ for  continuum simulations with $\mu_s=0.3$ and different values of  $\Delta \mu = 0.3$, $0.6$, $0.9$, $1.2$ and $1.5$ ($L=14.1d$). The inset graph shows the corresponding $\mu(I)$ dependence.}
\label{TuneDis2}
\end{figure}
\begin{figure}
\begin{minipage}{1.\linewidth} 
\centerline{\includegraphics[angle = - 0,width = 0.9\linewidth]{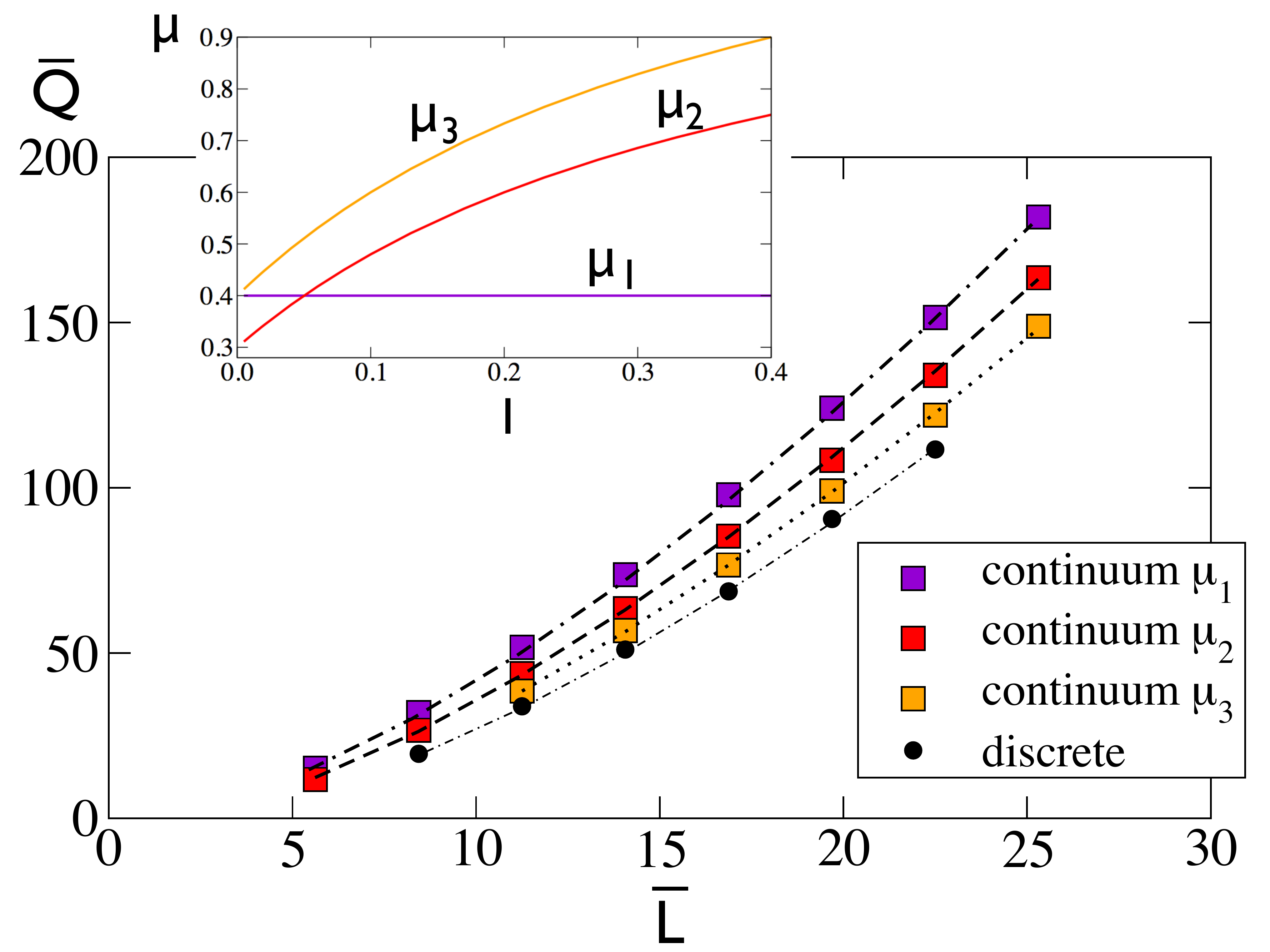}}
\end{minipage}
\caption{(Color on-line)  Normalized discharge rate $\bar{Q} = Q/\sqrt{g}d^{3/2}$ as a function of the normalized outlet size $\bar{L}= L/d$ for series of continuum simulations with different frictions laws (shown in the inset graph) and for discrete simulations. The dashed-dotted lines show the corresponding Beverloo fits.}
\label{TuneBev}
\end{figure}

\section{Tuning the rheological parameters}
\label{tunemui}

So far, no attempt was made to maximize the quantitative agreement between continuum and discrete approaches by adjusting appropriately the various rheological parameters. We recall the shape of the friction law adopted to approximate the continuum viscous behaviour: $\mu = \mu_s + {\Delta \mu} /(1+ I_0/I)$, where $\mu_s$ and $\Delta \mu$ set the value of the coefficient friction in the static and the highly dynamical limits. Two questions arise naturally from the comparison between discrete and continuum granular simulations. The first question is : Is the dependence of the friction $\mu$ on the inertial number $I$ a crucial ingredient for reproducing the granular phenomenology? The second question is: Is it possible to tune efficiently the different parameters to increase quantitative agreement?\\
To answer the first question, we perform a series of continuum simulations setting the parameter $\Delta \mu$ to zero: the dependence on the inertial number is suppressed. The static coefficient of friction $\mu_s$ is varied: $\mu_s=0.3$, $0.4$, $0.5$, $0.6$, $0.7$ and $0.8$. For each of these values, the corresponding continuum discharge is shown in Figure \ref{TuneDis1} for an outlet size $L=14.1d$. We observe that for moderate values of $\mu_s$, the discharge retains its linear shape, with a discharge rate diminishing with increasing $\mu_s$.  However, for larger values of $\mu_s$, the discharge loses its linear quality: the flow rate is no longer constant. This means that un-physically high values of the friction in the continuum model leads to a different behavior that would need precise characterization.  In any case, the agreement with the granular silo phenomenology is lost.\\
In a second set of simulations, we restore and investigate the dependence on the inertial number by varying $\Delta \mu$, alternatively set to $0.3$, $0.6$, $0.9$, $1.2$ and $1.5$, while $\mu_s$ is kept to a fixed value $0.3$.  The corresponding discharges are shown in Figure \ref{TuneDis2}. While the linearity is slightly compromised for $\Delta \mu=1.5$, we observe that tuning $\Delta \mu$ is  more efficient at slowing down the discharge rate than tuning $\mu_s$.  Hence, it appears that the dependence on the inertial number $I$, by including the shear-thickening properties of granular flows, allows for a more reliable description of  the discrete granular behavior than a constant friction model does. \\
 The effect of tuning the friction is illustrated in Figure~\ref{TuneBev} where the Beverloo scaling is plotted for continuum discharges with different parameters for the $\mu(I)$-flow-law. Increasing either or both $\mu_s$ and $\Delta \mu$, we obtain Beverloo scalings with smaller pre-factors $C$ and smaller effective outlet size $(L-k)$ (see scaling(\ref{eq:beverloo})). Eventually, although the constant $k$ which quantifies the steric constraints  due to the rigidity of the grains cannot be mimicked by the continuum flow, the pre-factor $C$ can be decreased to match  the discrete silo. Adjustment of the rheological parameters to achieve quantitative agreement between continuum and granular systems in terms of discharge is thus possible.  Understanding the origin of the discrepancies in the shape of the velocity profiles for low velocity discussed in sections \ref{compvel} and  \ref{comppre} would need however additional investigation.\\ 
 One may question the legitimacy of using large values of the coefficient of friction (between $0.6$ and $0.9$) in continuum models, while effective friction measured in granular systems is rarely beyond $0.5$ \cite{gdrmidi04}. In \cite{lagree11} for instance, quantitative agreement was achieved in the case of the collapse of granular columns using the same set of numerical parameters as applied here for both continuum  ($\mu_s = 0.32$, $\Delta \mu=0.28$ and $I_0=0.4$), and discrete ($\mu_c=0.5$, $e=0.5$) simulations. The failure to obtain quantitative agreement in the case of the silo for the same set of parameters suggests a possible dependence of the rheological properties on the geometrical flow configuration. It also question the relevance of  the $\mu(I)$ model for friction, valid for dense flow, in the outlet area, where the flow may become too dilute. %Finally, an aspect no treated in the present contribution, but of importance, is the influence of the size of the system. Indeed, the outlet size being typically  of a tenth of grain diameters, 

%On Fig \ref{TuneBev2}, we have $\mu_1(I) = 0.32 + \frac{0.28}{0.4/I+1}$, $\mu_2(I) = 0.32 + \frac{0.88}{0.4/I+1}$, $\mu_3(I) = 0.42 + \frac{0.98}{0.4/I+1}$, $\mu_4(I) = 0.32 + \frac{0.}{0.4/I+1}$.

%----  BOUNDARY CONDITiONS ----------------------
\begin{figure}
\begin{minipage}{1.\linewidth} 
\centerline{\includegraphics[angle = -90,width = 0.85\linewidth]{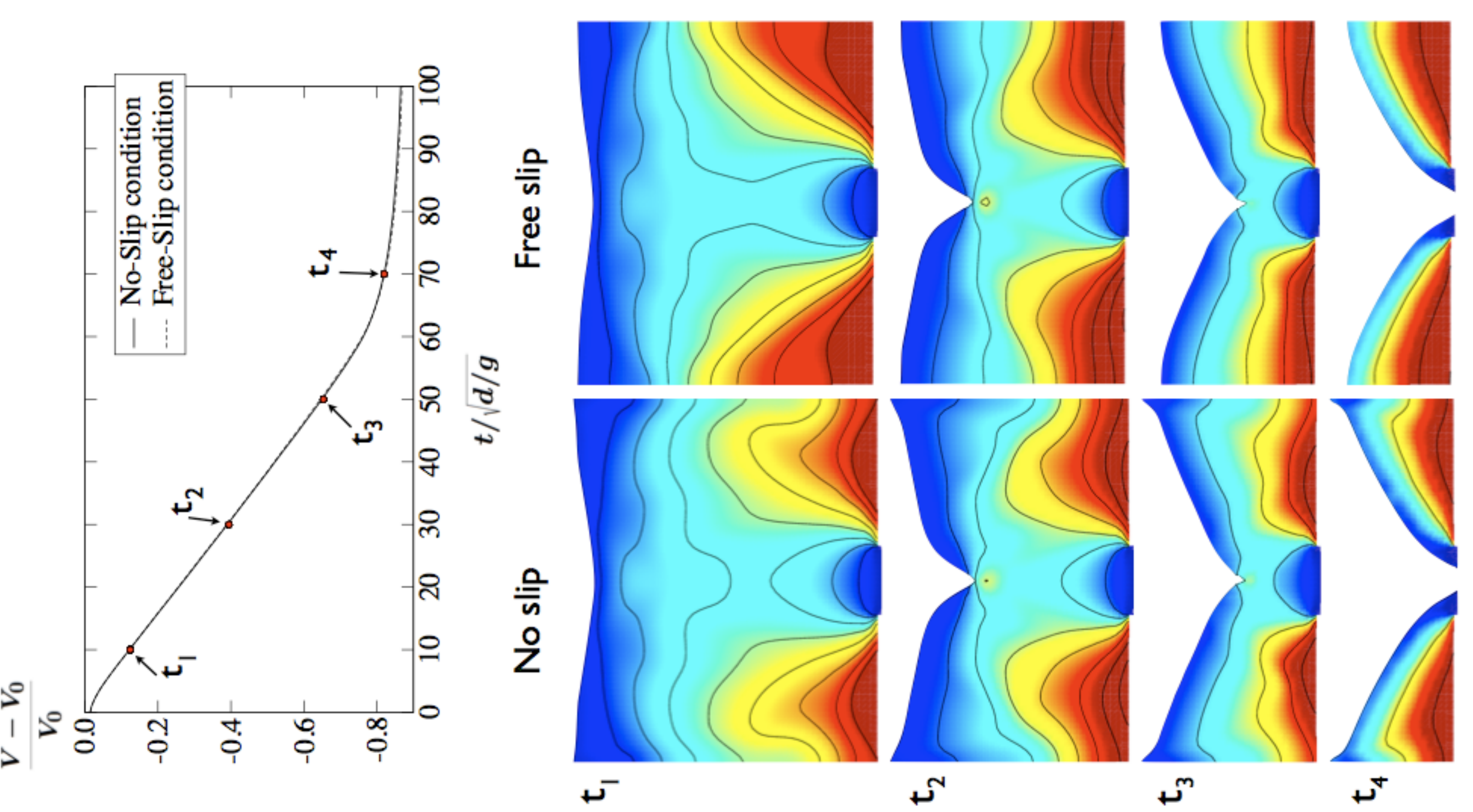}}
\caption{(Color On-line) Top graph:  Discharge of the continuum silo ($L= 14.1$) in the course of time for a no-slip condition and a free-slip condition at walls. Snapshots: map of the pressure at different instants $t_1$, $t_2$, $t_3$ and $t_4$ for no-slip (left) and free-slip (right) conditions.}
\label{BC}
\end{minipage}
\end{figure}

\section{Influence of the boundary conditions}
\label{sec:BC}

The boundary condition at the walls in all continuum simulations presented so far are no-slip boundary conditions. Although convenient, this may contradict the observation of slip velocities at rigid walls in a variety of experimental settings. In \cite{artoni12}, discrete simulation show that the slip velocity for dense granular flows obeys a Navier condition, with a dependence on the friction properties of the contacts. Without entering  this degree of description at this stage, one may nevertheless  speculate whether a free-slip velocity condition is more suited than a no-slip condition, and whether this change of boundary conditions would significantly affect the flow. \\
To clarify this aspect, we  perform a continuum discharge with a free-slip condition for the velocity at both side-walls, and compare it with the case of a no-slip condition. The volume left in the silo as a function of time for both cases  is shown in Figure \ref{BC}: the two evolutions  are virtually indistinguishable. In the course of the discharge, velocity and pressure profiles show marginal differences, and are essentially identical.  However, differences exist.  At the onset of the discharge,  the no-slip condition induces a lower pressure at the walls closer to the bottom than the free-slip condition does. This effect endures until the end of the discharge, although in a lesser extent; yet it has no consequence on the pressure field in the vicinity of the outlet and  in the flow area. More interestingly, the difference of boundary conditions affects the shape of the free surface close to the walls all through the discharge: the no-slip condition induces a convex tail of material adhering to the walls, when the free-slip condition has a nearly flat free surface in this area, more in accordance to the observation of granular systems. Far from the walls however, the shape of the free surface is identical for both boundary conditions. \\
We can thus conclude that the boundary conditions at the walls do not affect the essential features of the continuum silo discharge. This is expected since  the flow characteristics in a silo are dictated by the local condition created by the outlet, while velocities are small or zero far from that region.  It appears that the silo configuration is poorly suited to address in detail the relevance of free-slip, no-slip or mixed conditions at walls, and that other flow geometries would certainly give more information on the influence of boundary conditions when simulating continuum granular flows.

%------------------------------------------
\section{Conclusion}

Applying a continuum Navier-Stokes solver with the $\mu(I)$-flow-law implemented to model the viscous behavior of dense granular flows, we simulate the discharge of the granular silo  from the early stages of the discharge until complete release of the material.  Discrete simulations of the same system using the Contact Dynamics algorithm are performed to allow for systematic comparison between the two approaches.
Analysis of discharge rate, velocity field,  pressure field, and  surface and inner deformation is presented.
In a first step, we do not attempt to adjust the simulations rheological parameters to achieve quantitative agreement between continuum and discrete discharge rates, but focus on qualitative aspects. The Beverloo scaling is recovered in both cases. Analysing the shape of the velocity field at different locations in the flow, we find that continuum and discrete approaches show a good qualitative agreement in the areas of rapid flow, but tend to differ in the area  of slow shear, namely  near the outlet edges or in the bulk: the transition from a flowing state to a static one seems sharper in the continuum model. Focusing on the shape of the pressure profiles, the agreement also appears fairly good. Discrete and continuum simulations share the following features: a marked dip of pressure above the outlet,  the existence of two high pressure regions on one and the other side of the outlet, and a slight decrease close to the walls. 
Focusing on the shape of the free surface in the course of time, noticeable differences between the two approaches emerge:  the continuum system forms a dip above the outlet earlier then the discrete system, continuum slopes are slightly convex in shape when discrete slopes remain straight, and finally,  more material remains in the continuum silo than in the discrete one. However, the inner deformations for the two approaches are very similar. \\
Adjusting the rheological parameters of the continuum model to match quantitatively the discrete behavior reveals some potential problems. While the dependence on the inertial number, which is the key-ingredient of the friction law implemented in the  continuum model, is found to increase the ability of the latter to mimic discrete flows, the values of the continuum friction parameters needed to achieve quantitative correspondence between the two approaches are larger than the typical values measured in granular flows. In \cite{lagree11}, quantitative agreement was achieved in the case of the collapse of granular columns using the same set of numerical parameters as applied here for both continuum and discrete simulations. The failure to obtain quantitative agreement in the case of the silo for the same set of parameters suggests a possible dependence of the rheological properties on the geometrical flow configuration. Such known dependences include the width and height of a confined chute flow for instance \cite{jop05,pouliquen99}. This also suggests an effect of the size of the silo which would deserve further investigation, particularly in the case of small systems \cite{henann13}.
 The existence of non-local effects,  causing the friction to depend on the state of the system elsewhere rather than being a purely local property, may also possibly account for this discrepancy \cite{pouliquen09}. Their recent implementation in the case of steady flows leads to improved agreement between discrete and continuum granular systems, and allow for including systems size effects \cite{kamrin12,henann13}.  They are thus expected to allow for an improvement of the performances of the continuum model in the case of the silo configuration, and form a likely perspective of this work. At this stage however, we may conclude that the general ability of the continuum model  to reproduce the main features of the discharge of a granular silo is  very encouraging.

\section*{Acknowledgment}
The first author acknowledges financial support from the European Reseach Program FP7 IEF grant n$^\circ$~297843.

%Good results even for small apertures

%

\end{document}